\documentclass[aps,pre,reprint,amsmath,amssymb,superscriptaddress,showpacs,floatfix]{revtex4-1}
\usepackage{graphicx}
\usepackage{dcolumn}
\usepackage{bm}
\usepackage[colorlinks, allcolors=blue]{hyperref}
\usepackage[usenames, dvipsnames]{color}

\begin{document}

\title{Modeling the spin-Peierls transition of spin-$1/2$ chains with correlated states: $J_1-J_2$ model, CuGeO$_3$ and TTF-CuS$_4$C$_4$(CF$_3$)$_4$}

\author{Sudip Kumar Saha}
\affiliation{S. N. Bose National Centre for Basic Sciences, Block - JD, Sector - III, Salt Lake, Kolkata - 700106, India}

\author{Monalisa Singh Roy}
\affiliation{S. N. Bose National Centre for Basic Sciences, Block - JD, Sector - III, Salt Lake, Kolkata - 700106, India}

\author{Manoranjan Kumar}
\email{manoranjan.kumar@bose.res.in}
\affiliation{S. N. Bose National Centre for Basic Sciences, Block - JD, Sector - III, Salt Lake, Kolkata - 700106, India}

\author{Zolt\'an G. Soos}
\email{soos@princeton.edu}
\affiliation{Department of Chemistry, Princeton University, Princeton, New Jersey 08544, USA}

\date{\today}

\begin{abstract}

The spin-Peierls transition at $T_{SP}$ of spin-$1/2$ chains with isotropic exchange interactions has previously been 
modeled as correlated for $T > T_{SP}$ and mean field for $T < T_{SP}$. We use correlated states throughout in the $J_1-J_2$ model 
with antiferromagnetic exchange $J_1$ and $J_2 = \alpha J_1$ between first and second neighbors, respectively, and variable 
frustration $0 \leq \alpha \leq 0.50$. The thermodynamic limit is reached at high $T$ by exact diagonalization of short chains and 
at low $T$ by density matrix renormalization group calculations of progressively longer chains. In contrast to mean field results, 
correlated states of 1D models with linear spin-phonon coupling and a harmonic adiabatic lattice provide an internally consistent 
description in which the parameter $T_{SP}$ yields both the stiffness and the lattice dimerization $\delta(T)$. The relation 
between $T_{SP}$ and $\Delta(\delta,\alpha)$, the $T = 0$ gap induced by dimerization, depends strongly on $\alpha$ and deviates 
from the BCS gap relation that holds in uncorrelated spin chains. Correlated states account quantitatively for the magnetic susceptibility 
of TTF-CuS$_4$C$_4$(CF$_3$)$_4$ crystals ($J_1 = 79$ K, $\alpha = 0$, $T_{SP} = 12$ K) and CuGeO$_3$ crystals ($J_1 = 160$ K, $\alpha = 0.35$, 
	$T_{SP} = 14$ K). The same parameters describe the specific heat anomaly of CuGeO$_3$ and inelastic neutron scattering. Modeling the 
spin-Peierls transition with correlated states exploits the fact that $\delta(0)$ limits the range of spin correlations at $T = 0$ while $T > 0$ 
limits the range at $\delta= 0$.	
\end{abstract}
\maketitle
\section{\label{sec:intro}Introduction}
Jacobs et al.~\cite{jacob1976} identified the spin-Peierls transition at $T_{SP} = 12$ K in the organic crystal TTF-CuS$_4$C$_4$(CF$_3$)$_4$. The spin-$1/2$ 
chain at $T > T_{SP}$ has equally spaced cation radicals TTF$^+$ and is dimerized at lower $T$. They analyzed the magnetic susceptibility $\chi(T)$ using the 
linear Heisenberg antiferromagnet with equal exchange $J_1$ to both neighbors for $T > T_{SP}$ and alternating exchange $J_1(1 \pm \delta(T))$ in the dimerized phase. 
The $T$ dependence of $\delta(T)$ followed the BCS gap equation of superconductors. 
Subsequently, Hase et al.~\cite{haseprl1993, *haseprb1993} 
identified the inorganic spin-Peierls crystal CuGeO$_3$ with $T_{SP} = 14$ K based on spin-$1/2$ chains of Cu(II) ions. The magnetic susceptibility 
at $T > T_{SP}$ indicated~\cite{riera1995} exchange $J_2 = \alpha J_1$ with $\alpha= 0.35$ between second neighbors in addition to $J_1$. 
However, $\delta(T)$ did not follow BCS and extensive CuGeO$_3$ studies have been inconclusive~\cite{uchinokura2002} with respect to 
frustration $\alpha$. These prototypical spin-Peierls (SP) crystals have been analyzed with correlated states for $T > T_{SP}$ but only as 
uncorrelated or mean field for $T < T_{SP}$.

Spin-$1/2$ chains have been long studied theoretically as simple 1D systems with two states, $\alpha$ and $\beta$, per site. The linear Heisenberg 
antiferromagnet (HAF) is the $\alpha= 0$ limit of the $J_1-J_2$ model, Eq.~\ref{eq:j1j2} below. The HAF may well be the best characterized many-body system, 
and the $J_1-J_2$ model also has an extensive literature. 

The electronic problem for SP transitions is to obtain the thermodynamic limit of the free energy per 
site $A(T,\delta)$ at temperature $T$ and dimerization $\delta$. In reduced ($J_1 = 1$) units, we have
\begin{equation}
	A(T,\delta)=-T \ln Q (T,\delta).
\label{eq:freeen}
\end{equation}
The thermodynamic limit is known for free fermions but not for correlated systems such as the HAF or the $J_1-J_2$ model.
SP modeling has consequently been approximate and subject to revision due to computational advances. In particular, we show below that
$\delta(T)$ for the HAF does $not$ follow BCS. 



We model both transitions with a recent method that combines
exact diagonalization (ED) 
of short chains with density matrix renormalization group (DMRG) calculations of progressively longer chains~\cite{sudip19}. The 
premise is that the full spectrum $\lbrace E(\delta,N) \rbrace$ of large systems is never needed. Since $T$ limits the range of 
spin correlations, ED is sufficient once the system size exceeds the correlation length. Bonner-Fisher results~\cite{bonner64} 
to $N = 12$ were used~\cite{jacob1976} for $\chi(T)$ of TTF$^+$ chains at $T > T_{SP}$. ED to $N = 24$ is now accessible. DMRG for 
larger $N$ yields the spectrum $\lbrace E(\delta,N) \rbrace$ up to some cutoff $E_C(\delta,N)$, thereby extending thermodynamics to lower $T$. 
The hybrid approach is particularly well suited for SP systems because dimerization opens a gap that limits spin correlations at $T = 0$. 

The driving force for dimerization is the partial derivative $\partial A(T,\delta)/\partial \delta$ that is opposed by the 
lattice. The simplest lattice model is used in conventional approaches~\cite{su1980, beni1972, soosfreo2002} to the Peierls or 
SP instability: the coupling is linear, the potential energy $\delta^2/2\varepsilon_d$ per site is harmonic, and the stiffness $1/\varepsilon_d$ 
is independent of $T$. The equilibrium dimerization is
\begin{equation}
	\frac{\delta(T)}{\varepsilon_d}=-\left(\frac{\partial A(T,\delta)}{\partial \delta} \right)_{\delta(T)}.
\label{eq:potentialen}
\end{equation}
At $T = 0$, $A(0,\delta) = E_0(\delta)$ is the ground state energy per site. DMRG returns the derivative $E_0^\prime(\delta,N)$ of large systems 
and the extrapolated limit $E_0^\prime(\delta)$. Dimerization decreases and vanishes at $T_{SP}$, where $1/\varepsilon_d = -A^{\prime \prime} (T_{SP},0) $. 
In principle, the observed $T_{SP}$ is 
the model parameter that specifies both the stiffness and $\delta(T)$. To emphasize the point, we refer to the equilibrium susceptibility 
as $\chi(T,T_{SP})$ over the entire range. Moreover, the driving force is a property of the electronic system that is balanced by 
whatever model is adopted for the lattice.

The equilibrium dimerization is explicitly known for free fermions; $\delta(T)$ for a half-filled tight-binding band is given by
\begin{equation}
\begin{aligned}	
	& \frac{1}{\varepsilon_d}=\frac{8}{\pi} \int_0^{\pi/2} dk \frac{\sin^2 k}{\varepsilon\left( k,\delta(T) \right)} \tanh \frac{\varepsilon(k,\delta(T))}{2T}, \\ 
	& \varepsilon(k,\delta)=2\sqrt{\cos^2 k +\delta^2 \sin^2 k}.
\end{aligned}
	\label{eq:dimerization_tight_binding}
\end{equation}
The stiffness is half as large for spinless fermions, which corresponds to the XY spin-$1/2$ chain. The band gap opens as $2\varepsilon(\pi/2,\delta) = 4\delta$ and
$\delta(0)$ goes as $\exp(-1/\varepsilon_d)$ in the weak coupling limit. The spinless fermion representation of the HAF has interactions
between first neighbors. The HAF is correlated. Although not exact, the HAF gap opens~\cite{barnes99} as $\delta^{3/4}$ based on diverse
numerical studies collected in Ref.~\onlinecite{johnston2000}. The DMRG exponent in the range $0.001 \leq \delta \leq 0.10$ is~\cite{mkumar2007}
$0.7475 \pm 0.0075$.



\begin{figure}
\includegraphics[width=\columnwidth]{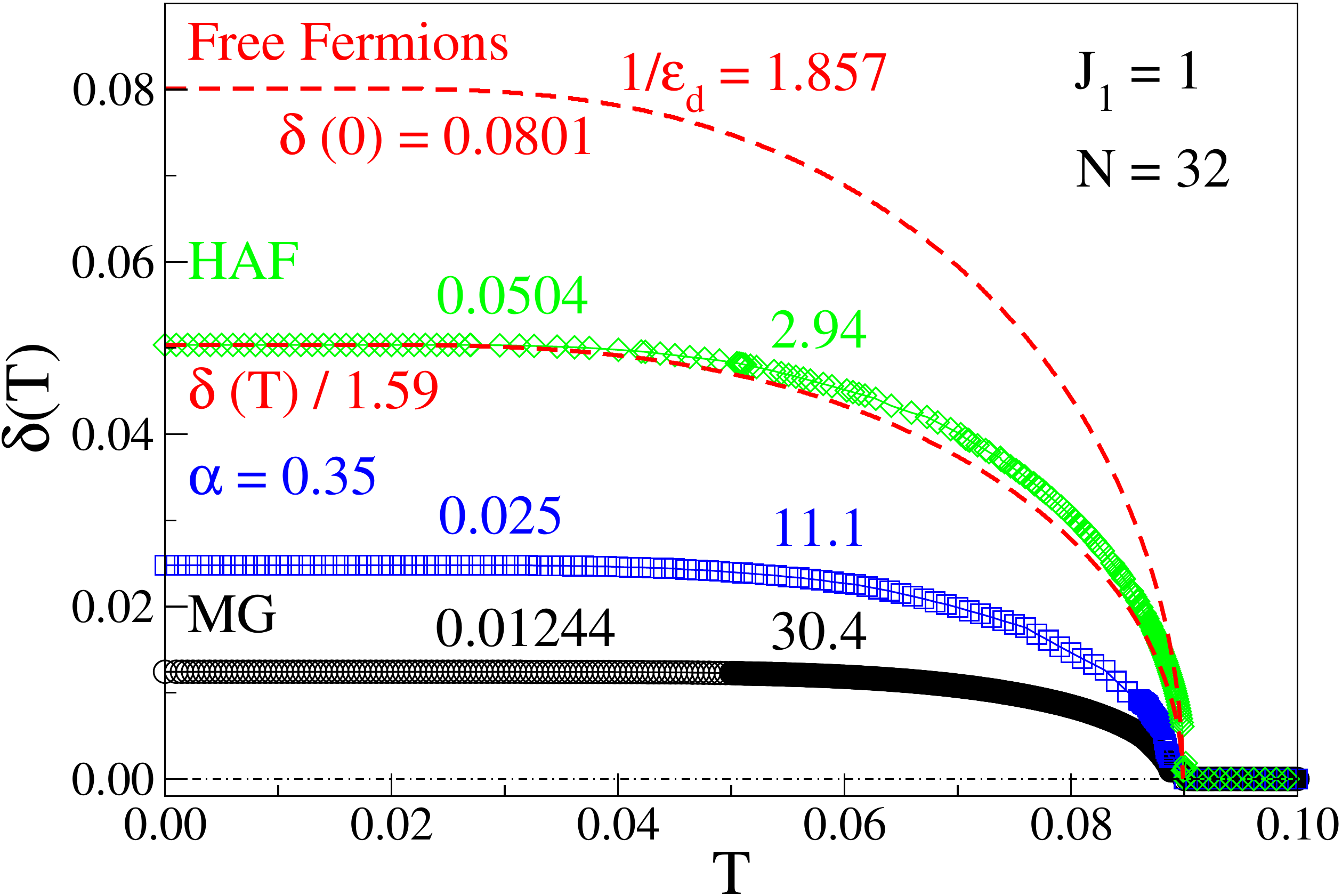}
\caption{\label{fig1}
	Equilibrium dimerization $\delta(T)$ of spin chains with $T_{SP} = 0.09$ leading to stiffness 
$1/\varepsilon_d$ and $\delta(0)$ in Eq.~\ref{eq:potentialen}. The exact free fermion curve is Eq.~\ref{eq:dimerization_tight_binding} with $4/\pi$ 
	instead of $8/\pi$. The HAF ($\alpha = 0$), $\alpha= 0.35$ and MG ($\alpha= 0.50$) curves are based on Eq.~\ref{eq:j1j2} with $N = 32$ spins. The 
HAF dimerization is close to the fermion $\delta(T)/1.59$. Note the large variation of $1/\varepsilon_d$ and $\delta(0)$ in chains with equal $T_{SP}$.} 
\end{figure}

To illustrate correlations and frustration, we show in Fig.~\ref{fig1} the dimerization of spin chains with $T_{SP} = 0.09$ (or $0.09 J_1$). 
The fermion curve is Eq.~\ref{eq:dimerization_tight_binding} 
with $4/\pi$ instead of $8/\pi$; the band gap $4\delta(0) = 3.55 T_{SP}$ is within $1\%$ of the BCS gap relation. The other curves are $A^\prime(T,\delta,N)$ 
for $J_1-J_2$ models with $N = 32$ in Eq.~\ref{eq:j1j2}, periodic boundary conditions and $\alpha = 0$ (HAF), $0.35$ and $0.50$ (MG). 
The fermion $\delta(T)$ scaled by $1/1.59$ is the dashed line through the HAF points; the scaled $T$ dependence is nearly BCS.
The stiffness increases by an order of magnitude from the HAF to MG while $\delta(0)$ decreases by a factor of four and $\delta(T)$ clearly 
deviates from free fermions.

%

We analyze SP transitions of the $J_1-J_2$ model with frustration $0 \leq \alpha \leq 0.50$. Under some conditions, numerical advances have made accessible 
the thermodynamic limit of correlated states of 1D systems.
The influential but approximate HAF analysis~\cite{bray1983,jacob1976} of TTF-CuS$_4$C$_4$(CF$_3$)$_4$ was widely thought to
apply to the larger data set made possible by sizeable CuGeO$_3$ crystals. But CuGeO$_3$ turned out to be different and has largely resisted modeling. Correlated states provide a consistent description of both SP transitions.

The paper is organized as follows. Section~\ref{sec2} presents the calculation of $A^\prime(T,\delta)$ in $J_1-J_2$ models with frustration $\alpha$
and the criterion for the thermodynamic limit.
We model in Section~\ref{sec3} the magnetic susceptibility $\chi(T,T_{SP})$ of TTF-CuS$_4$C$_4$(CF$_3$)$_4$ with two parameters, $J_1 = 79$ K 
and $T_{SP} = 12$ K. The CuGeO$_3$ parameters $J_1 = 160$ K, $\alpha= 0.35$ and $T_{SP} = 14$ K account for both $\chi(T,T_{SP})$ and the specific 
heat anomaly, $C(T,T_{SP})$. In Section~\ref{sec4} we discuss the CuGeO$_3$ excitations probed by inelastic neutron scattering, 
not modeled previously, that give an independent determination of $J_1$. We also study the Majumdar-Ghosh (MG) point~\cite{ckm69b}, $\alpha= 0.50$, 
where the exact ground state is known. Aspects and limitations of 1D models are mentioned in the Discussion.

\section{\label{sec2} Dimerized $J_1-J_2$ model}
The $J_1-J_2$ model has isotropic exchange interactions $J_1$, $J_2 = \alpha J_1$ between first and second neighbors of a 
regular ($\delta= 0$) spin-$1/2$ chain. The dimerized model has alternating $J_1(1 \pm \delta)$ along the chain. We consider 
finite chains with $N = 4n$ spins, periodic boundary conditions and $J_1 = 1$ as the unit of energy. The electronic Hamiltonian is  
\begin{equation}
        H(\delta,\alpha) = \sum_{r} \left( 1+\delta (-1)^r  \right)   \vec{S}_r \cdot \vec{S}_{r+1} + \alpha \sum_{r} \vec{S}_{r} \cdot \vec{S}_{r+2}.
\label{eq:j1j2}
\end{equation}
The HAF is the special case $\alpha=\delta=0$. The ground state of $H(0,\alpha)$ is nondegenerate for $0 \leq \alpha \leq \alpha_c = 0.2411$, the 
quantum critical point~\cite{nomura1992} that separates a gapless phase from the gapped dimer phase with a doubly degenerate ground state. The exact $\delta = 0$ 
ground state is known at $\alpha= 0.50$, the MG point~\cite{ckm69b}, that marks the onset of an incommensurate phase. 
Finite $\delta$ breaks inversion symmetry at sites and increases the singlet-triplet gap $\Delta(\delta,\alpha)$ but does not change 
the length in systems with periodic boundary conditions. The analysis does not depend on the index $\alpha$ which is suppressed below.

We consider the equilibrium Eq.~\ref{eq:potentialen} with increasing system size to obtain the thermodynamic limit at finite $T$ 
and then evaluate $\delta(T)$ in models with $T_{SP} > T$. The free energy per spin of finite chains is
\begin{equation}
	A(T,\delta,N)=-T N^{-1} \ln Q (T,\delta,N).
\label{eq:freeen_finite}
\end{equation}
The Boltzmann sum in $Q(T,\delta,N)$ is over the $2^N$ spin states with energies $E_r (\delta,N)$. Exact diagonalization (ED) 
yields the full spectrum of short chains. The equilibrium dimerization requires the partial derivative that we approximate as
\begin{equation}
        A^\prime(T,\delta,N)\approx   \frac {A(T,\delta+\varepsilon,N)-A(T,\delta-\varepsilon,N)}{2\varepsilon}.
\label{eq:freederivative}
\end{equation}
The numerator is accurate to three decimal places for $\varepsilon = 0.001$. We find that the size dependence 
of $A^\prime$ is considerably weaker than that of $A$, presumably due to cancellations in the numerator.

The hybrid ED/DMRG method~\cite{sudip19} follows the size dependence of the quantity of interest, here the driving force
$-A^\prime(T,\delta,N)$. Since $T$ reduces the 
range of spin correlations, ED up to $N = 24$ for $\delta= 0$ or $N = 20$ for $\delta> 0$ returns the thermodynamic limit at high $T$. 
DMRG with periodic boundary conditions~\cite{ddpbc2016} is then used to obtain the lowest few thousand states of larger systems. The 
spectrum $E_r(\delta,N) \leq E_C(\delta,N)$ up to a cutoff defines a truncated partition function $Q_C(T,\delta,N)$ and hence a truncated 
entropy per site, $S_C(T,\delta,N)$. Finite size gaps reduce $S_C(T,\delta,N)$ compared to the actual entropy at low $T$ while truncation 
reduces it at high $T$. Since $S_C(T,\delta,N)/T$ converges from below with increasing $N$, its maximum at $T^\prime(\delta,N)$ is the best 
choice for a given cutoff $E_C(\delta,N)$. The cutoff is increased until $T^\prime(\delta,N)$ is independent or almost independent of $E_C$. 
The thermodynamic limit of $A^\prime(T^\prime,\delta)$ at $T^\prime(\delta,N)$ is approximated by DMRG at system size $N$.

\begin{figure}
\includegraphics[width=\columnwidth]{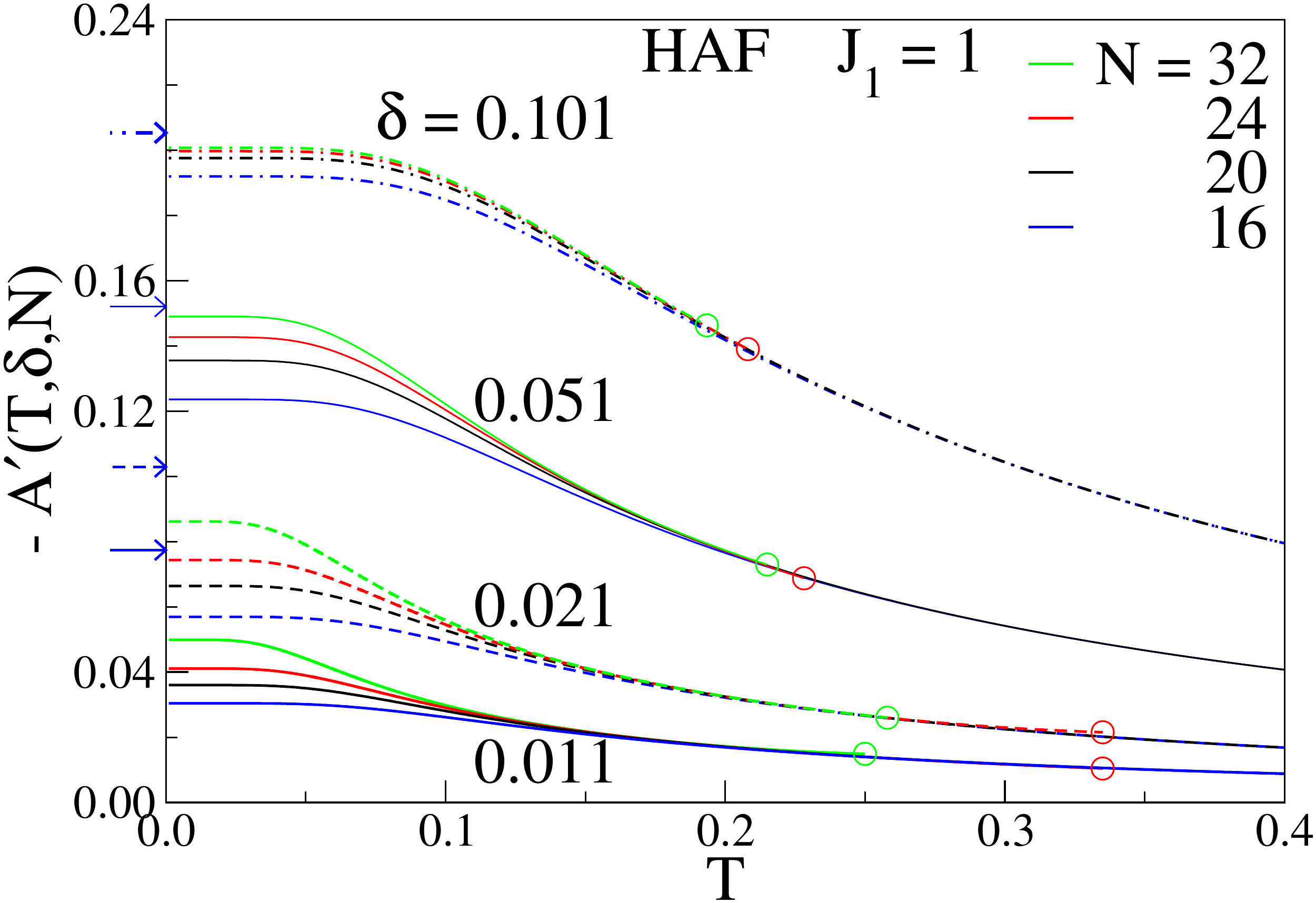}
\caption{\label{fig2}
Driving force for dimerization, $-A^\prime(T,\delta,N)$, of HAF chains with $N$ spins and $\alpha = 0$ in 
Eq.~\ref{eq:j1j2}. $N = 16$ and $20$ are exact. DMRG for $N = 24$ and $32$ is shown up to $T^\prime(\delta,N)$, shown as open 
	circles, the maximum of 
$S_C(T,\delta,N)/T$ discussed in the text. Arrows at $T = 0$ are thermodynamic limits that increase with $\delta$.
	}
\end{figure}

Fig.~\ref{fig2} illustrates the $T$ dependence of $-A^\prime(T,\delta,N)$ of the HAF. As expected for any $\alpha$, $-A^\prime$ decreases with $T$ and increases $N$ to the thermodynamic limit. The $N = 16$ and $20$ lines are exact. DMRG results 
for $N > 20$ extend to the points $T^\prime(\delta,N)$, the maxima of $S_C(T,\delta,N)/T$ that are shown as open circles. 
Finite size gaps are evident 
around $T \sim \delta \sim 0$ where $-A^\prime(T,\delta,N)$ is constant. Arrows indicate the $T = 0$ intercepts, $-E_0^\prime(\delta,N)$, 
that are obtained by extrapolation of ground-state DMRG calculations~\cite{mkumar2007} at constant $\delta$. Since $\delta$  opens a magnetic gap in the 
infinite chain, the size dependence decreases as seen at $\delta= 0.101$. Convergence to the thermodynamic limit is found by $T \sim 0.15$, The general criterion based on $T^\prime(\delta,N)$ is evidently conservative for $A^\prime(T,\delta,N)$, which is seen to converge at lower $T$.

\begin{figure}[b]
\includegraphics[width=\columnwidth]{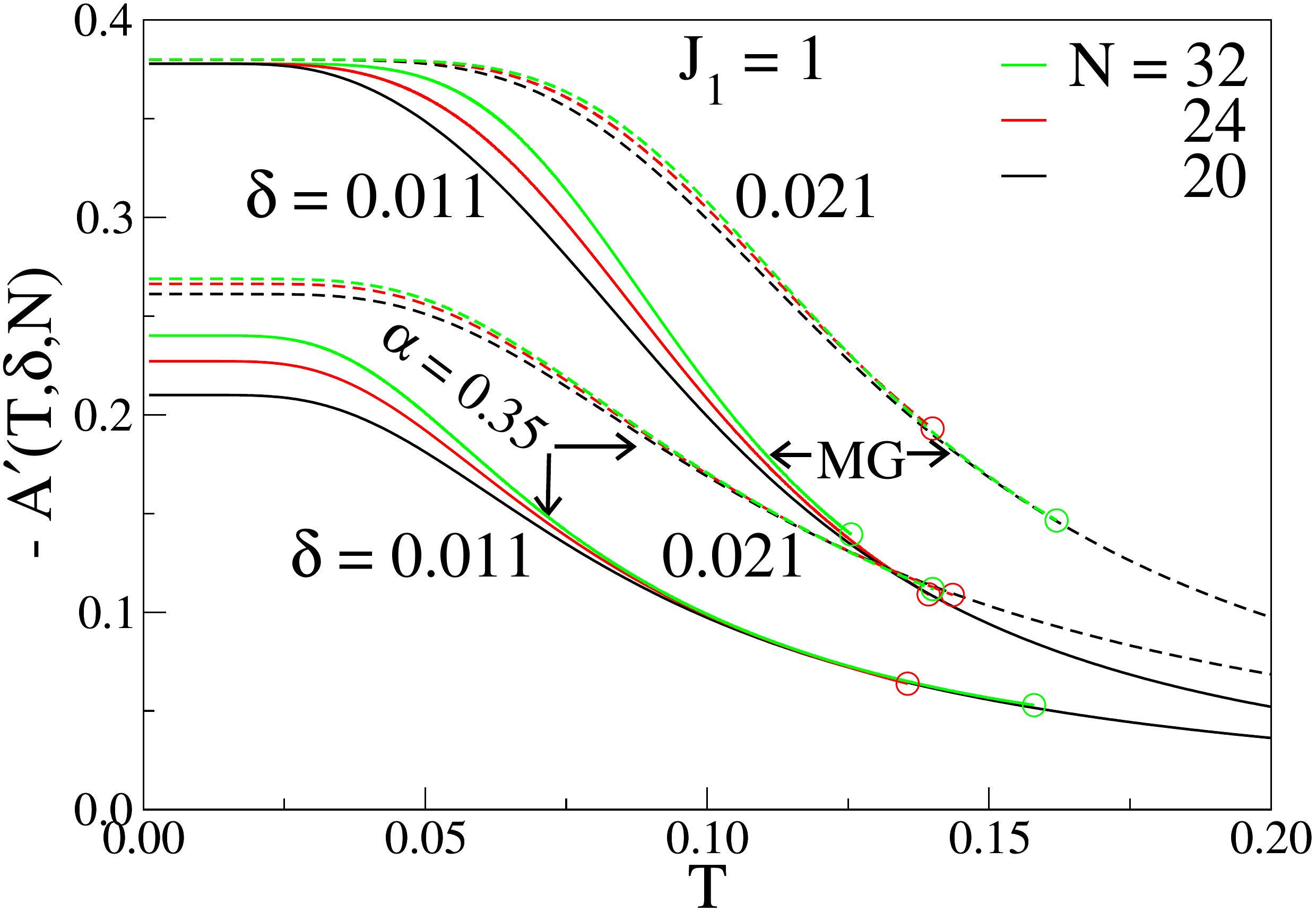}
	\caption{\label{fig3} Same as Fig.~\ref{fig2} for frustration $\alpha = 0.35$ and MG ($\alpha=0.50$) in Eq.~\ref{eq:j1j2}.}
\end{figure}

The size dependence of $A^\prime(T,\delta,N)$ in the dimer phase is shown in Fig.~\ref{fig3} for $\alpha= 0.35$ and $0.50$ in Eq.~\ref{eq:j1j2}. 
The MG ground states are the two Kekul\'e VB diagrams with singlet pairing either between all sites $2r$, $2r - 1$ or all sites $2r$, $2r + 1$. The energy per site 
is $-3/8$ for even $N$ in Eq.~\ref{eq:j1j2} and $A^\prime(0,\delta,N) = - 3/8$ is exact~\cite{saha2019spinpeierls} to order $\delta$. 
The thermodynamic limit is reached by $T \sim 0.13$ for $\alpha= 0.50$. 
The size dependence at $\alpha= 0.35$ is intermediate. The ground state is degenerate in the thermodynamic limit but not for finite $N$. 
The $-A^\prime(0,0,N)$ intercept decreases with $N$ to $B(0.35) = 0.078$ in the thermodynamic limit, where $B(\alpha)$ is the amplitude 
of the bond order wave~\cite{mkumarbndord2010}. The size dependence again decreases with $\delta$.  

Fig.~\ref{fig2} and Fig.~\ref{fig3} indicate how $A^\prime(T,\delta,N)$ approaches the thermodynamic limit. The convergence depends on the 
model and the largest system $N_m$, 
\begin{equation}
	A^\prime(T,\delta,N_m) \rightarrow A^\prime(T,\delta),  \qquad  T>T^\prime(\delta,N_m).
\label{eq:freethermo}
\end{equation}
$T^\prime(\delta,N_m)$ is the maximum of $S_C(\delta,T,N_m)/T$ of the largest system considered. We have 
performed DMRG calculations up to $N \sim 100$, but smaller $N$ may be sufficient and convergence at $\delta=0$ typically also
holds for $\delta >0$. 
The system size is eventually limited~\cite{sudip19} by the numerical accuracy of the dense energy spectrum, 
which is of course model dependent. Although the mathematically interesting $A^\prime(T,0)$ at 
$T \sim 0$ is out of reach, modeling SP transitions merely requires $T_{SP} > T^\prime(0,N_m)$. The equilibrium 
Eq.~\ref{eq:potentialen} then gives $\delta(T)$ in the thermodynamic limit.

\begin{figure}
\includegraphics[width=\columnwidth]{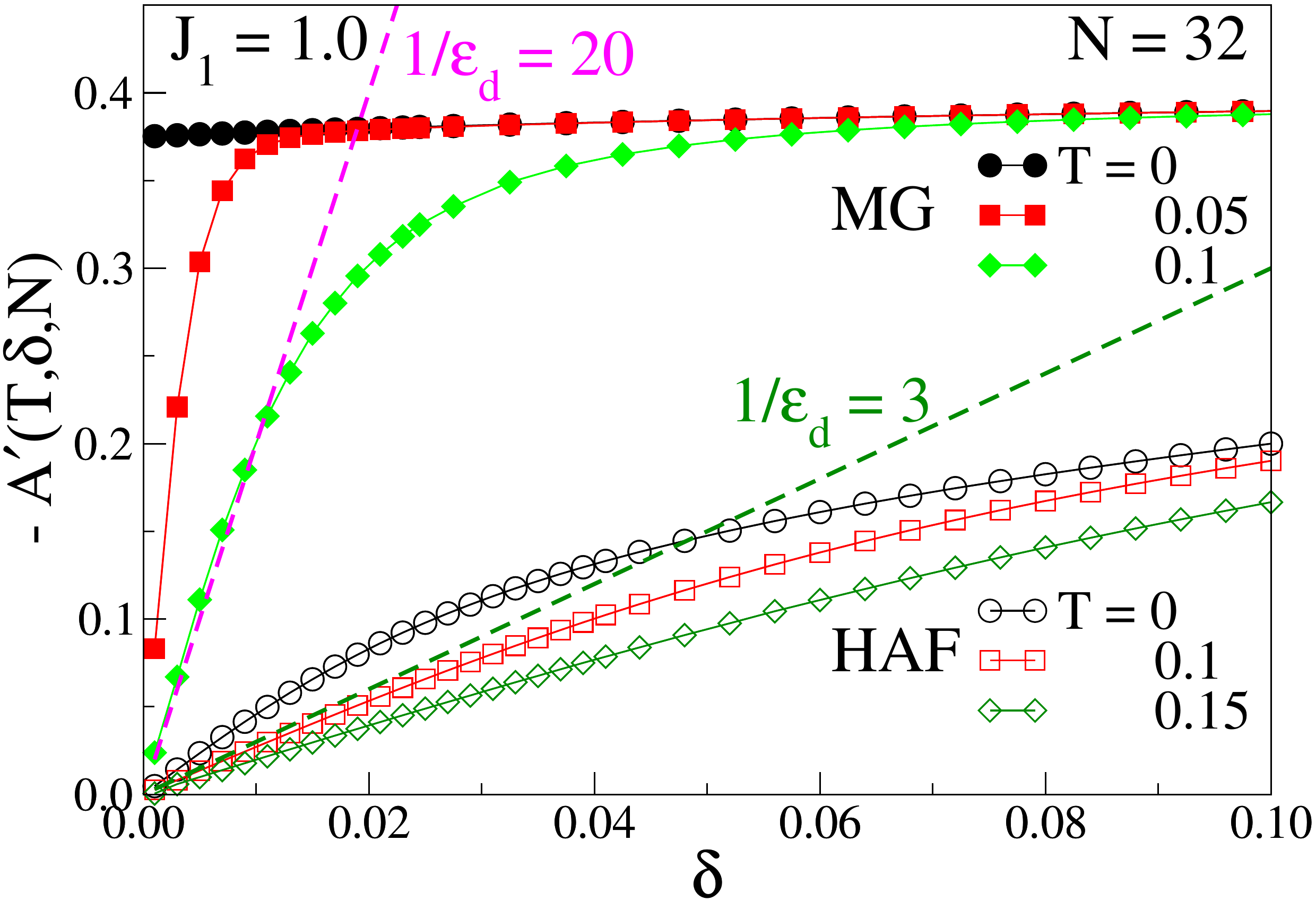}
\caption{\label{fig4}
The driving force $-A^\prime(T,\delta,N)$ at reduced $T$ of MG and HAF chains with $N = 32$ 
and $\alpha= 0.50$ and $0$ in Eq.~\ref{eq:j1j2}. The dashed lines are $\delta/\varepsilon_d$ and crossing points 
	are solutions $\delta(T)$ to Eq.~\ref{eq:potentialen}. 
	The HAF and MG curves are lower and upper bounds for frustration $0 \leq \alpha \leq 0.50$.
        }
\end{figure}

Fig. ~\ref{fig4} shows $A^\prime(T,\delta,N)$ 
vs. $\delta$ for models with $N = 32$ and $\alpha = 0$ (HAF) or $0.50$ (MG). These curves lower and upper bounds of $-A^\prime(T,\delta,N)$ for $J_1-J_2$ models
with $0 \leq \alpha \leq 0.50$.
The $\delta = 0$ 
intercept at $T = 0$ decreases from $3/8$ at $\alpha= 0.50$ to zero at $\alpha_c = 0.2411$ where~\cite{cross79,mkumar2007}  $E_0^\prime(\delta) 
= -0.62\delta^{0.33}$. The HAF result~\cite{mkumar2007} is $E_0^\prime(\delta) = –0.56\delta^{0.44}$. The graphical solutions $\delta(T,\alpha)$ of Eq.~\ref{eq:potentialen}
are the intersections in Fig.~\ref{fig4} of $A^\prime(T,\delta,N)$ with dashed lines $\delta/\varepsilon_d$ at the indicated stiffness. The chains are unconditionally unstable for finite
$\varepsilon_d$ since $E_0^\prime(\delta)$ is finite at $\delta = 0$ for $\alpha > \alpha_c$ while $E_0^{\prime \prime}(\delta)$ diverges at $\delta = 0$ for $\alpha < \alpha_c$.
The $E_0^\prime(\delta)$ cusp at $\delta= 0$ in the dimer phase leads to the flatter two $\delta(T)$ curves~\cite{saha2019spinpeierls}
in Fig.~\ref{fig1}.

%

We conclude that the thermodynamic limit $A^\prime(T,\delta)$ can be reached in finite chains at $T = 0$ 
when $\delta > 0$ or at $\delta= 0$ when $T > 0$. The $T > T_{SP}$ range is more accessible numerically for large $T_{SP}/J_1$  
that in turn generates large $\delta(0)$. The relation between $T_{SP}$ and $\delta(0)$ is strongly model dependent as seen in Fig.\ref{fig1}.


%
%

\section{\label{sec3} Magnetic susceptibility and specific heat}
\begin{figure}
\includegraphics[width=\columnwidth]{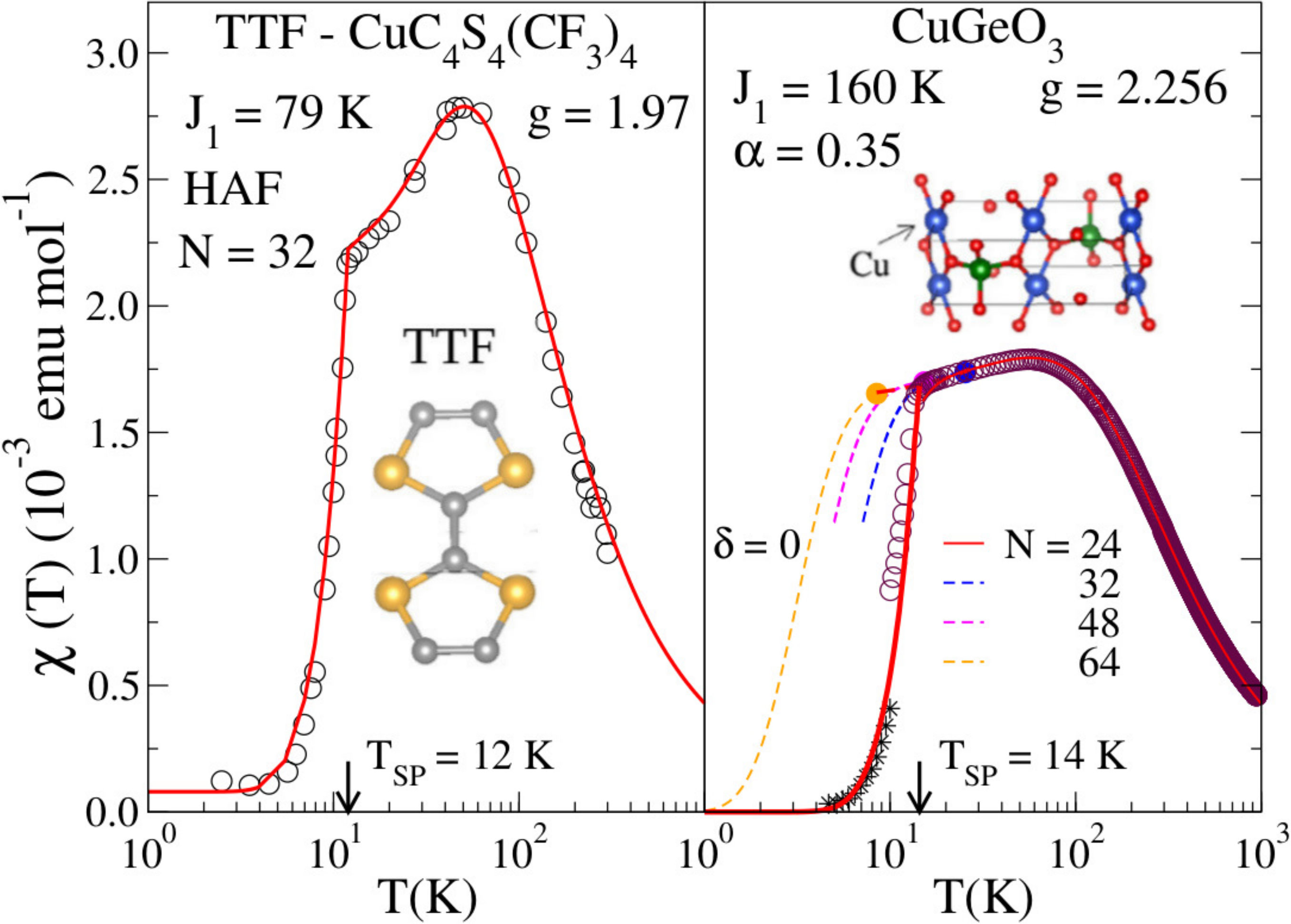}
\caption{\label{fig5} 
Absolute molar magnetic susceptibility: TTF-CuS$_4$C$_4$(CF$_3$)$_4$ data from Fig. 5 of Ref.~\onlinecite{jacob1976} or 
Fig. 10 of Ref.~\onlinecite{bray1983}; 
CuGeO$_3$ data from Ref.~\onlinecite{haseprl1993, *haseprb1993} to $10$ K and Ref.~\onlinecite{fabricius1998} for $T > 10$ K. Fits are discussed in the text. The $\delta= 0$ lines are DMRG up to $T^\prime(0,N)$ 
shown as filled circles. 
	}
\end{figure}

A sudden decrease of the molar magnetic susceptibility $\chi(T)$ at $T < T_{SP}$ is a direct manifestation of an SP transition. 
Fig.~\ref{fig5} shows published data for~\cite{jacob1976} TTF-CuS$_4$C$_4$(CF$_3$)$_4$ 
and~\cite{fabricius1998,haseprl1993} CuGeO$_3$ on a log scale that emphasizes low $T$. The excellent TTF$^+$ fit shown in Fig.~\ref{fig5} of Ref.~\onlinecite{jacob1976} or Fig. 10 of Ref.~\onlinecite{bray1983} is based
%
on the HAF with $J_1 = 77$ K and $g = 1.97$ for $T > T_{SP} = 12$ K. The $g$ value is within the range given by electron spin resonance (esr) 
with the applied magnetic field along the $c$ axis. The $T \sim 0$ limit is shown as slightly positive ($\sim 0.08 \times 10^{-3}$ emu/mol). 

We included 
this $T$-independent contribution in the correlated fit shown with $J_1 = 79$ K, $T_{SP} = 12$ K and $g = 1.97$. The $\delta= 0$ curve above $T_{SP}$ 
is ED for $N = 24$ and DMRG for $N = 32$ with $\alpha= 0$ in Eq.~\ref{eq:j1j2}. In the dimerized phase, we calculated $\chi(T,T_{SP})$ for $N = 32$ at 
the equilibrium $\delta(T)$ given by Eq.~\ref{eq:potentialen}. The correlated fit is equally quantitative. It has one fewer parameter and is internally consistent: $T_{SP}$ 
and $J_1$ determine both the stiffness $1/\varepsilon_d = 1.96$ and $\delta(0) = 0.103$. The previous $\chi(T,\delta(T))$ was based~\cite{jacob1976} on a mean field $\delta(T)$ 
for the $T$ dependence and required an adjustable $\delta(0) = 0.126$ that, as noted,~\cite{jacob1976} leads to $T_{SP} = 9$ rather than $12$ K.

The range of spin correlations is reduced at low $T$ by substantial dimerization $\delta(0) = 0.10$. 
The thermodynamic limit is reached in relatively short chains that are now amenable to quantitative analysis. Correlated states clarify the SP 
transition of TTF-CuS$_4$C$_4$(CF$_3$)$_4$. Contrary to long held expectations, the HAF dimerization $\delta(T)$ does \textit{not} follow free fermions or BCS.

The $\chi(T)$ data for CuGeO$_3$ are from Ref.~\onlinecite{haseprl1993, *haseprb1993} up to $10$ K ($T_{SP} = 14$ K) and from Ref.~\onlinecite{fabricius1998} 
from $10$ to $950$ K ($T_{SP} = 14.3$ K), kindly provided in digital form by Professor Lorenz. There is a mismatch at $10$ K. The range ($\sim 0.5$ K) 
of reported $T_{SP}$ reflect variations of growth conditions that are discussed in Ref.~\onlinecite{hidaka1997}. We retained the previous 
parameters~\cite{fabricius1998} based on ED for $N = 18$ and the $\chi(T)$ maximum at $T = 56$ K: $J_1 = 160$ K, frustration $\alpha= 0.35$ in Eq.~\ref{eq:j1j2}, $g = 2.256$ from esr. 
The $\delta= 0$ fit is quantitative for $T > T_{SP} = 14$ K ($0.09 J_1$). The points $T^\prime(N)$ on the $\delta= 0$ curve are the $S_C(T,0,N)/T$
maxima of truncated calculations at system size $N$. The resulting $\chi(T,T_{SP})$ for $T < T_{SP}$ is consistent with the available data and 
corresponds to $\delta(0) = 0.025$. We extend~\cite{fabricius1998} or improve~\cite{castilla1995,bouzerar1999} previous $T > T_{SP}$ fits.

\begin{figure}
\includegraphics[width=\columnwidth]{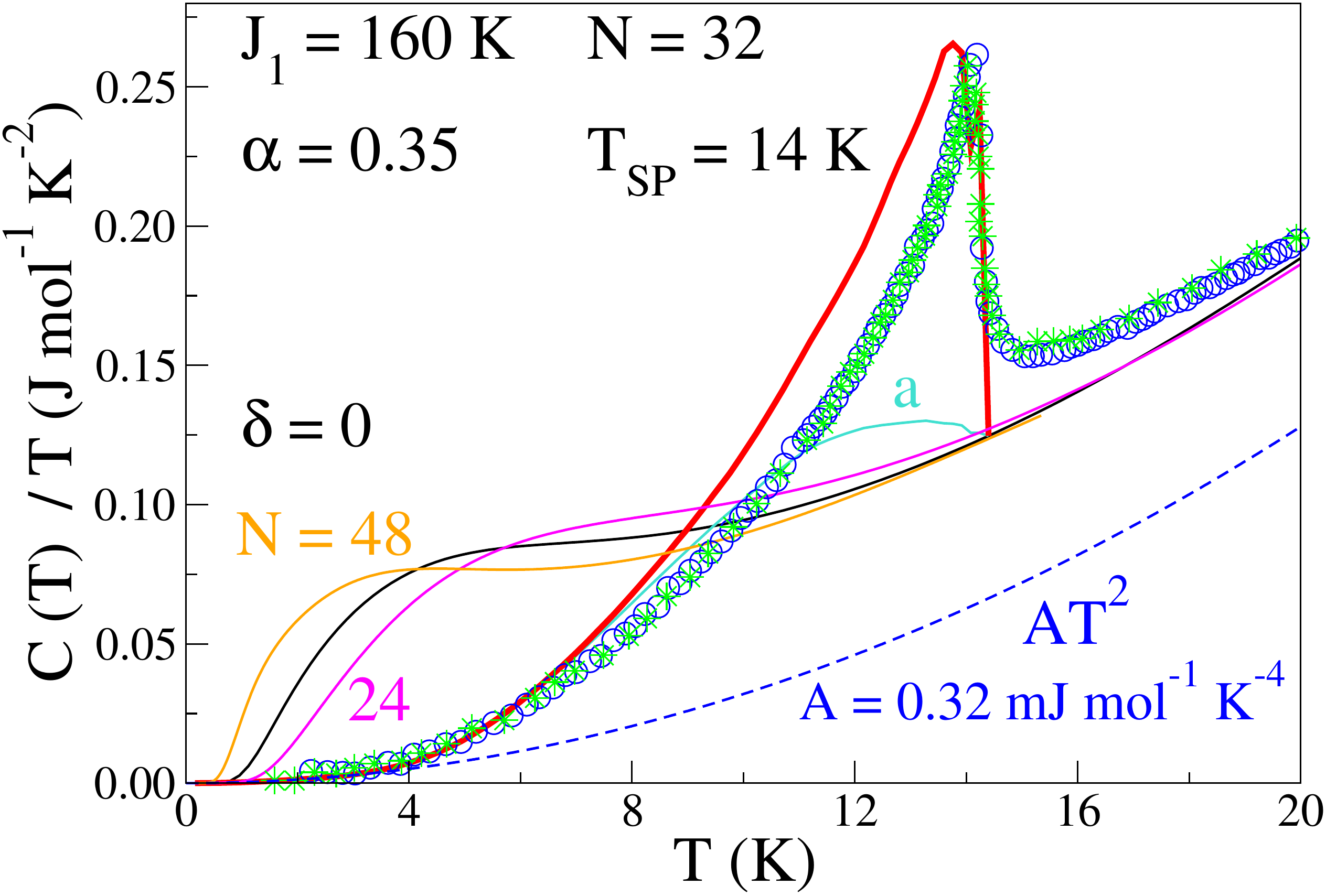}
\caption{\label{fig6}
Molar specific heat $C(T)$ of CuGeO$_3$ shown as the entropy derivative $S^\prime = C/T$: blue circles from Ref.~\onlinecite{lorenz1996},  green stars 
and A from Ref.~\onlinecite{liu1995}. The calculated $C(T,0)/T$ curves are ED for $N = 24$, DMRG for $N = 32$ and $48$. The equilibrium $C(T,T_{SP})/T$ is the bold red line,  Eq.~\ref{eq:specific_heat}, whose first term is labeled (a). 
	}
\end{figure}

Sizeable single crystals of CuGeO$_3$ made possible other measurements. The specific heat $C(T)$ to $20$ K is shown in Fig.~\ref{fig6} as the entropy derivative 
$S^\prime = C/T$ in Refs.~\onlinecite{lorenz1996,liu1995}. The dashed line 
is the reported lattice (Debye) contribution~\cite{liu1995}, 
$AT^2$, with $A= 0.32$ mJ/mol K$^4$. The specific heat has not been modeled aside from the initial exponential increase with $T$. The anomaly 
is sharper and better resolved than in small TTF-CuS$_4$C$_4$(CF$_3$)$_4$ crystals~\cite{wei1977}.

The equilibrium $C(T,T_{SP})$ has two contributions~\cite{saha2019spinpeierls} below $T_{SP}$,
\begin{equation}
\begin{aligned}
	& C(T,T_{SP})=C(T,\delta(T)) + \\
        & \qquad \frac {\partial \delta} {\partial T} \left[ \left( \frac {\partial E(T,\delta)} {\partial \delta} \right)_T +\frac {\delta(T)} {\varepsilon_d} \right].
\label{eq:specific_heat}
\end{aligned}
\end{equation}
$E(T,\delta)$ is the internal energy per site, $-(\partial \ln Q(\beta,\delta)/\partial \beta)$  with $\beta= 1/k_B T$. The first 
term is evaluated at $\delta(T)$ along the equilibrium $\alpha= 0.35$ line in Fig. 1, which can be fit quantitatively as
\begin{equation}
\frac {\delta(T)} {\delta(0)} = \left(1-\left(\frac {T} {T_{SP}} \right)^{a} \right)^{b}, \qquad  T \leq T_{SP}
\label{eq:equi_dimer_T}
\end{equation}
with $a = 5.29$ and $b = 0.689$. We used Eq.~\ref{eq:equi_dimer_T} to evaluate $\partial \delta/\partial T$. The calculated 
$C(T,T_{SP})/T$ is the bold red line shown in Fig.~\ref{fig6}. The low-$T$ behavior of $\delta= 0$ chains is a finite size effect. Since gaps initially 
decrease $C(T,N)/T$, entropy conservation requires increased $C(T,N)/T$ before converging from above to the thermodynamic limit. 
The $N = 48$ and $24$ gaps are smaller and larger, respectively, than $N = 32$, which is in the thermodynamic limit for $T > 12$ K.

The $C(T,\delta(T))/T$ part of Eq.~\ref{eq:specific_heat} is the curve labeled (a) in Fig.~\ref{fig6}. The $\partial \delta(T)/\delta T$ 
derivative  is mainly responsible for the sharp anomaly. The area under $C(T,T_{SP})/T$ up to $T_{SP}$ is within $5\%$ of the accurately known $\delta= 0$ area. 
The adiabatic and mean field approximations for the lattice enforce $\delta= 0$ for 
$T > T_{SP}$; this general problem for any transition has long been recognized.
The agreement between theory and experiment by $20$ K implies equal area under the measured, dimerized and $\delta= 0$ curves in Fig.~\ref{fig6}. Lattice fluctuations observed above $T_{SP}$ must be offset by reduced $C/T$ below $T_{SP}$. Overall, the anomaly is fit rather well considering these approximations.

\section{\label{sec4} Inelastic neutron scattering}

\begin{figure}[t]
\includegraphics[width=\columnwidth]{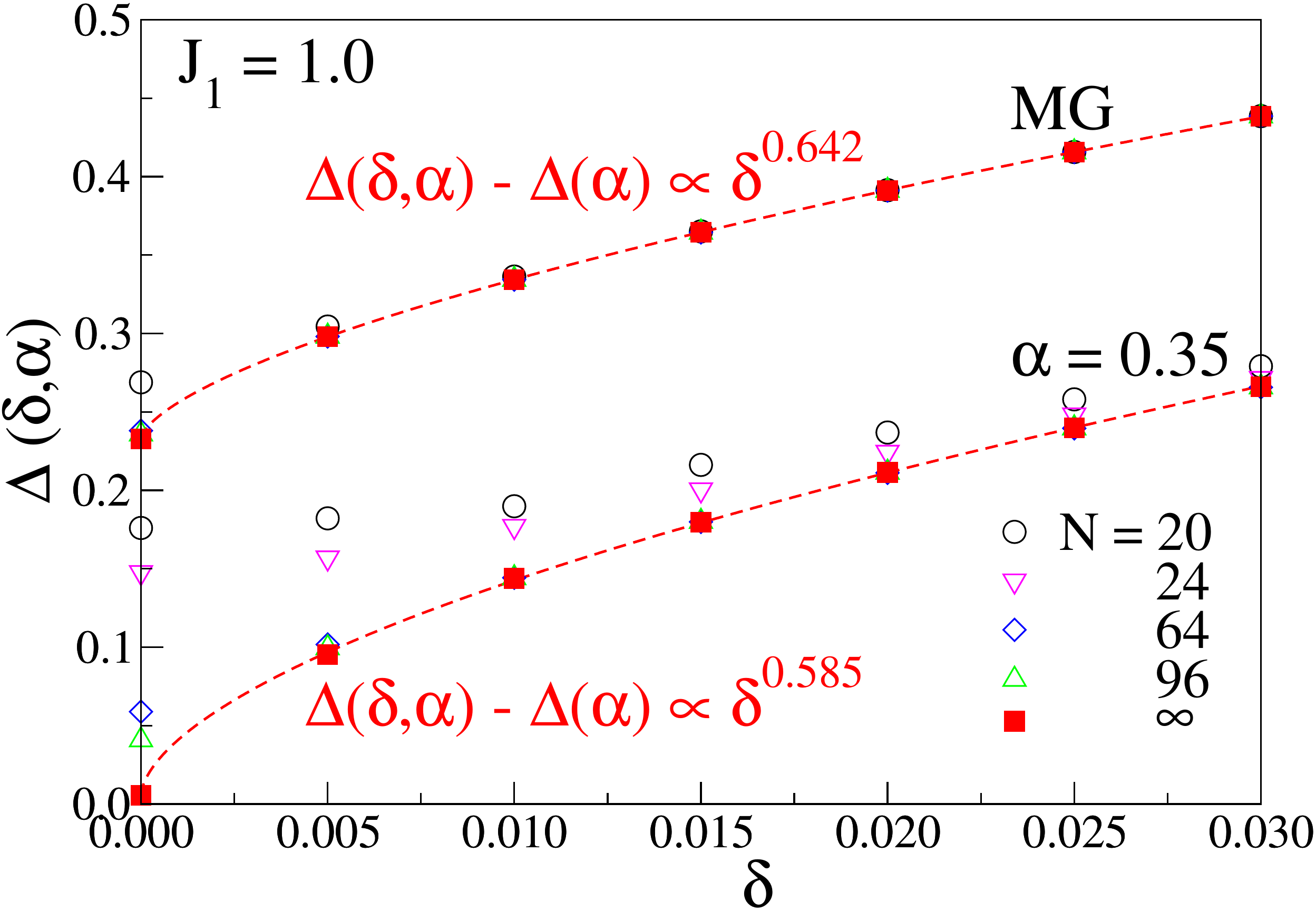}
\caption{\label{fig7}
Scaled singlet-triplet gap $\Delta(T,\delta,N)$ at $T = 0$ of $N$-spin chains in Eq.~\ref{eq:j1j2} with dimerization $\delta$ and frustration 
$\alpha= 0.35$ and $0.50$. The lines are $1/N$ extrapolations of DMRG to $N = 96$. Note the rapid convergence at the MG point.
	}
\end{figure}
%


Dimerization opens a gap $\Delta(\delta,\alpha)$ in gapless spin chains or increases the gap in gapped chains. 
The gap is from the singlet ($S = 0$) ground state to the lowest energy triplet ($S = 1$). 
The opening of the HAF gap~\cite{barnes99,johnston2000} $\Delta(\delta,0)$ or of $\Delta(\delta,\alpha_c)$ at the critical point~\cite{cross79,mkumar2007} 
has been extensively discussed using field theory and numerical methods; $\Delta(0,\alpha)$ is finite in the dimer phase, exponentially small 
just above $\alpha_c$ and substantial at $\alpha = 0.50$. We obtained the thermodynamic limit of $T = 0$ gaps in Fig.~\ref{fig7} by extrapolation
of DMRG calculations up to $N = 96$. As expected, size convergence is rapid for $\delta > 0.01$. The gap opens as


\begin{equation}
\Delta(\delta,\alpha)=\Delta(\alpha)+D\delta^\gamma,
\label{eq:stgap}
\end{equation}
with $\Delta = 0.0053$, $D = 2.03$ and $\gamma=0.585$ for $\alpha= 0.35$. The $T$ dependence is given by $\delta(T)$. The large MG gap
is $\Delta(0,0.5) = 0.233$.

%
\begin{figure}
\includegraphics[width=\columnwidth]{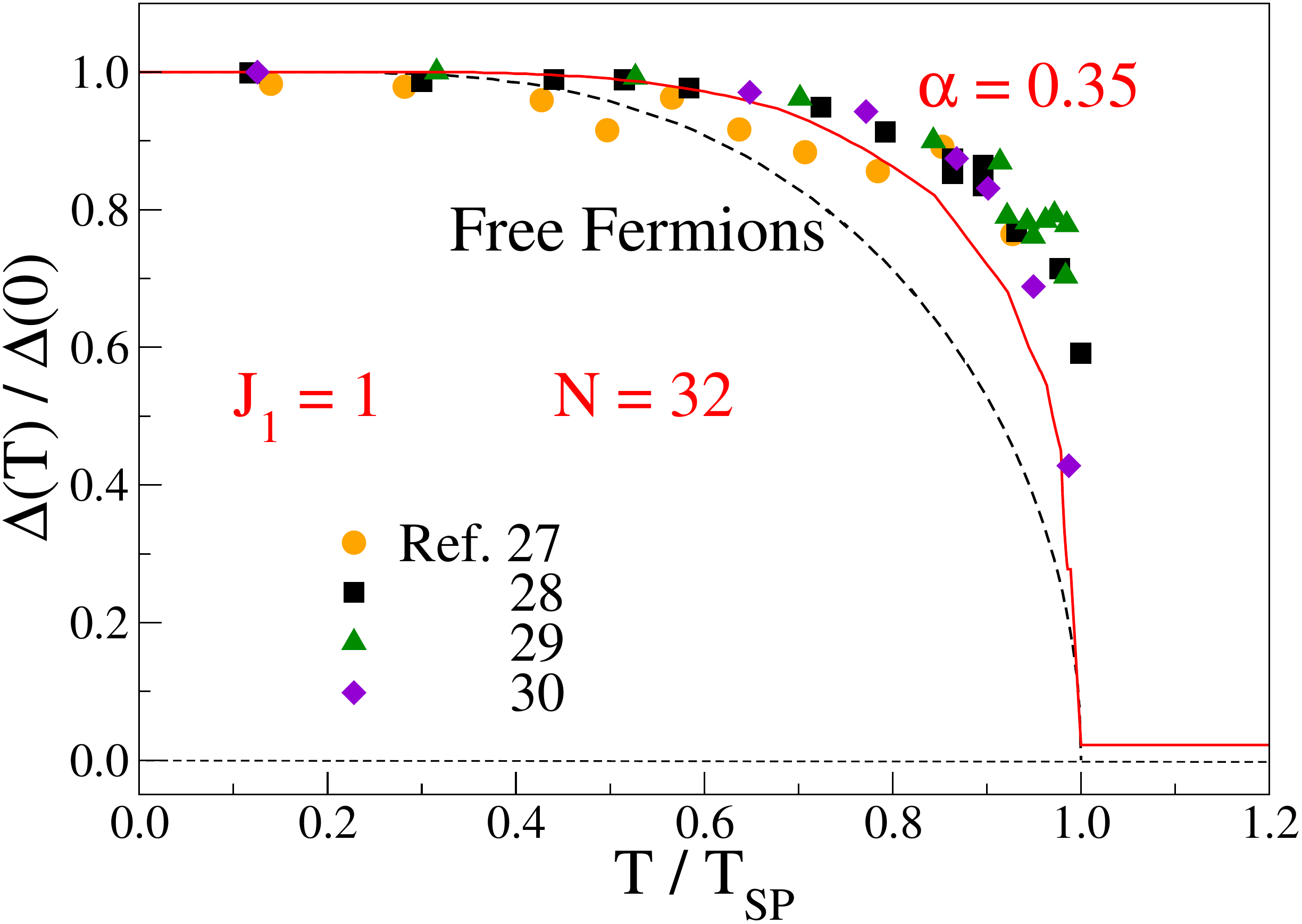}
\caption{\label{fig8}
Scaled singlet-triplet gap $\Delta(T)/\Delta(0)$ vs. $T/T_{SP}$.
	The solid line is $\Delta(\delta(T),\alpha)/\Delta(\delta(0),\alpha)$ with $\alpha= 0.35$ in Eq.~\ref{eq:stgap}; the dashed line is $\delta(T)/\delta(0)$ for free
	fermions in Fig~\ref{fig1}. The symbols are inelastic neutron data from Refs.~\onlinecite{nishi1994,lussier1996,martin1996,regnault96}.
	}
\end{figure}

Inelastic neutron scattering (INS) at $T = 0$ is exclusively to triplets in models with isotropic exchange. 
Fig.~\ref{fig8} shows the scaled gap $\Delta(T)/\Delta(0)$ vs. $T/T_{SP}$. The solid line is the calculated $\Delta(\delta(T),0.35)$ in Eq.~\ref{eq:stgap} with $\delta(T)$ 
in Eq.~\ref{eq:equi_dimer_T} or the $\alpha= 0.35$ curve in Fig.~\ref{fig1}. The gap $\Delta J_1 = 0.85$ K is almost an order of magnitude 
below INS resolution and scales to $0.022$ for $T > T_{SP}$ and $\delta(0) = 0.025$. The dashed line is the gap ratio $\delta(T)/\delta(0)$ 
for free fermions in Fig.~\ref{fig1}. INS studies of CuGeO$_3$ crystals in Refs.~\cite{nishi1994,lussier1996,martin1996,regnault96} have reported the $T$ 
dependence of the singlet-triplet gap $\Delta(T)$. 
(The organic crystals are unsuitably small~\cite{regnault96}.) Large deviations from the free fermions or BCS were unexpected and unexplained. 
Correlated states are consistent with these data, and how quantitatively remains to be seen.

The calculation of the INS spectrum is straightforward in finite systems with periodic boundary conditions.
Triplets $\vert T_n(q)\rangle$ at $E_n(q)$ relative to the singlet ground state $\vert G\rangle$ are required. At $T = 0$, the INS intensity 
$M_n(q)$ for energy transfer $\omega = E_n(q)$ and momentum transfer $q$ is~\cite{muller1981}
\begin{equation}
\begin{aligned}
   & M_n(q)=2\pi \vert \langle T_n(q)\vert S^z_q\vert G \rangle \vert ^2, \\
	& S^z_q = (4n)^{-\frac{1}{2}} \sum_r e^{iqr} S^z_r.
\end{aligned}
\label{eq:matelem}
\end{equation}
The $x$ or $y$ components of $S_q$ also yield $M_n(q)$. The lowest triplet of $H(\delta,\alpha)$ is $E_1(\pi)=\Delta(\delta,\alpha)$
The INS intensity at finite 
$T$ is the thermal average~\cite{muller1981} of Eq.~\ref{eq:matelem} over excited states as well as $\vert G\rangle$. 
The static structure factor at $T = 0$ is given by ground-state spin correlation functions  
\begin{equation}
	S(q)=\langle G\vert S^z_{-q} S^z_q \vert G \rangle.
\label{eq:static_str_fac}
\end{equation}
The total INS intensity per spin is $\pi/2$ for chains with a singlet ground state. The thermal average of $S(T,q)$ in Eq.~\ref{eq:static_str_fac}
is far less tedious since it only requires the $T$ dependence of $N/2$ correlation functions. 

\begin{figure}
\includegraphics[width=\columnwidth]{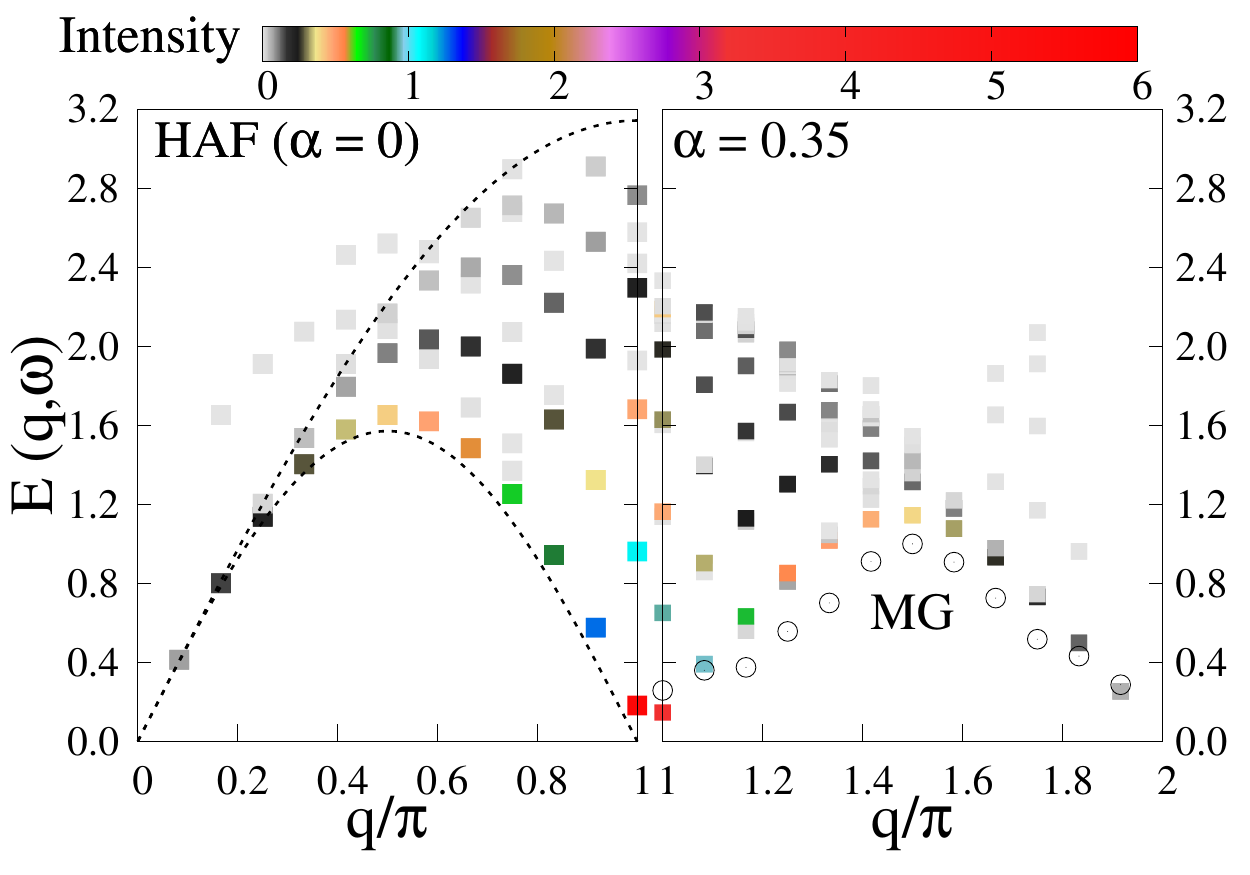}
\caption{\label{fig9}
Exact triplet excitations $E_n(q)$ at wave vector $q$ in $24$-spin chains: left panel, $\alpha = 0$ (HAF); right panel, $\alpha= 0.35$. 
	The color coding is the intensity $M_n(q) > 0.001$ in Eq.~\ref{eq:matelem}. The lowest triplets $E_1(q)$ for $\alpha= 0.50$ (MG) are the open circles on the right. 
The dashed lines are the spinon boundaries, Eq.~\ref{eq:spinonrange}.
	}
\end{figure}

The Bethe ansatz~\cite{bethe31,*hulthen38} has provided the exact ground state of the HAF and the so-called Class $C$ states with $S > 0$ 
that can be solved exactly. Faddeev and Takhtajan~\cite{faddeev81} obtained the double spinon continuum in the thermodynamic limits. 
The lower and upper boundaries at wave vector $q$ are
\begin{equation}
\begin{aligned}
	& \varepsilon_1 (q)=\frac{\pi}{2} \sin q; \qquad 0 \leq q \leq \pi  \\
	& \varepsilon_2 (q)=\pi \sin \frac{q}{2}.
\end{aligned}
\label{eq:spinonrange}
\end{equation}
Each state is four-fold degenerate, two $S = 1/2$ spinons forming a triplet or a singlet. The boundaries up to $q = \pi$ are the dashed 
lines in the HAF panel of Fig.~\ref{fig9}. The spectrum is symmetric about $q = \pi$. The singlet-triplet gap $\varepsilon_1(q)$ was found 
earlier by des Cloizeaux and Pearson~\cite{pearson1962}. 

The almost quantitative calculation of intensities $M_n(q)$ in the thermodynamic limit has recently been 
achieved~\cite{karbach97}.
Mourigal et al.~\cite{mourigal2013} have confirmed theory in detail on a Cu(II) spin chain with $J_1 = 2.93$ K; the INS analysis 
in Fig. 1d of Ref.~\onlinecite{mourigal2013} was carried out at finite $T$ using both two and four-spinon calculations. $S(q,\omega)$ is continuous in the thermodynamic limit.
Fig. 1d is color coded according to intensity and impressive agreement between theory and experiment is shown, as in Fig.~\ref{fig9}, side by side with
$0 \leq q \leq \pi$ and $\pi \leq q \leq 2\pi$.


%

When total spin is conserved, a system of $N = 4n$ spins has $3(4n)!/[(2n -1)!(2n + 2)!]$ triplets out of which only $n(2n + 1)$ are in Class $C$~\cite{bethe31,*hulthen38} and 
have excitation energy between $\varepsilon_2(q)$ and $\varepsilon_1(q)$ in the thermodynamic limit. The $S(q,\omega)$ spectra 
in the HAF panel of Fig.~\ref{fig9} are $\alpha=\delta=0$ and $N = 24$. The color coding is according to the 
intensity $M_n(q) > 0.001$. There are a few triplets not in Class $C$, but $99.4\% $ of the total intensity is between the dashed 
lines. The discrete $S(q,\omega)$ spectra are close to the thermodynamic limit for both excitations and intensities.

The $S(q,\omega)$ spectra in the $\alpha = 0.35$ panel of Fig.~\ref{fig9} are for $\delta= 0$, $N = 24$ and color coded according to $M_n(q) > 0.001$. ED returns the full spectrum. The special feature of the 
triplets shown is greater intensity than over $500,000$ other triplets. The triplets account for $99.5\%$ of the total intensity and, 
again with a few outliers, resemble the spinons in the left panel.

Frustration decreases the dispersion $E_1(q)$ of the lowest triplet, as shown by open circles in the MG ($\alpha=0.50$) curve. 
The HAF triplets $E_1(q,N)$ are slightly above $\varepsilon_1(q)$, 
which at $q = \pi$ is entirely due to finite size. The $q = \pi$, $\alpha= 0.35$ gap is mainly due to finite size while the $\alpha= 0.50$ gap 
is close the thermodynamic limit of $0.233$ in Fig.~\ref{fig7}. 
The HAF dispersion has 
previously been used to infer $J_1 = 2E_1/\pi$ from the measured $E_1(\pi/2)$.

Arai et al.~\cite{arai1996} reported the $S(q,\omega)$ spectrum of CuGeO$_3$ and interpreted it using the HAF while also pointing out differences. 
At $10$ K, the observed $S(q,\omega)$ intensity peaks at $\pi/2$ and $3\pi/2$ are at $16$ meV ($186$ K). The peaks for $\alpha= 0.35$, $N = 24$ in 
Fig.~\ref{fig9} are at reduced energy $E_1(\pi/2) = 1.14$, or $E_1 = 182$ K for $J_1 = 160$ K. The agreement is well within the combined accuracy. The $N = 16$, $20$ and $24$ 
gaps extrapolated as $1/N$ return $E_1 = 1.1$ in the thermodynamic limit. The weak size dependence is typical of large gaps. 
The upper limit of the INS spectrum extends~\cite{arai1996} to $32$ meV at $q=\pi$ at both $10$ and $50$ K.
The calculated $T = 0$ spectrum with appreciable $M_n(\pi)$ also extends to $\omega \sim 2E_1 (\pi/2) = 2.28$.

The calculations in Fig.~\ref{fig9} approximate the unknown $S(q,\omega)$ at $\alpha= 0.35$ in the same sense that $N = 24$ approximates the HAF spectrum. 
$S(q,\omega)$ at $q = \pi/2$ or $3\pi/2$ of CuGeO$_3$ has a noticeably narrower~\cite{arai1996} energy spread than the spinon spread 
$\varepsilon_2(\pi/2) - \varepsilon_1(\pi/2)$. The correlated states in Fig.~\ref{fig9} capture this narrowing at $\alpha= 0.35$ compared to $\alpha= 0$. 
Indeed, the width is entirely suppressed at $\alpha= 0.50$ where INS at $q = \pi/2$ or $3\pi/2$ is a $\delta$-function at $E = J_1$. This exact result for a 
triplet, not reported previously, is derived in the Appendix.

The INS data in Fig.~\ref{fig10}, upper panel, for the static structure factor $S(T,q)$ of CuGeO$_3$ is rescaled Fig. 2 of Ref.~\onlinecite{arai1996}. 
Large differences from HAF were noted~\cite{arai1996}. The dashed line is the exact {$S(q) = (1 - \cos q)/4$} at $T = \delta= 0$, $\alpha= 0.50$, where 
finite size simply leads to discrete $q$. Although $S(q)$ depends on $\alpha$ and $\delta$ the area $\pi/2$ under $S(q)$ does not. The $10$ K data are almost 
as broad as the MG curve before considering the resolution in $q$.
\begin{figure}
\includegraphics[width=\columnwidth]{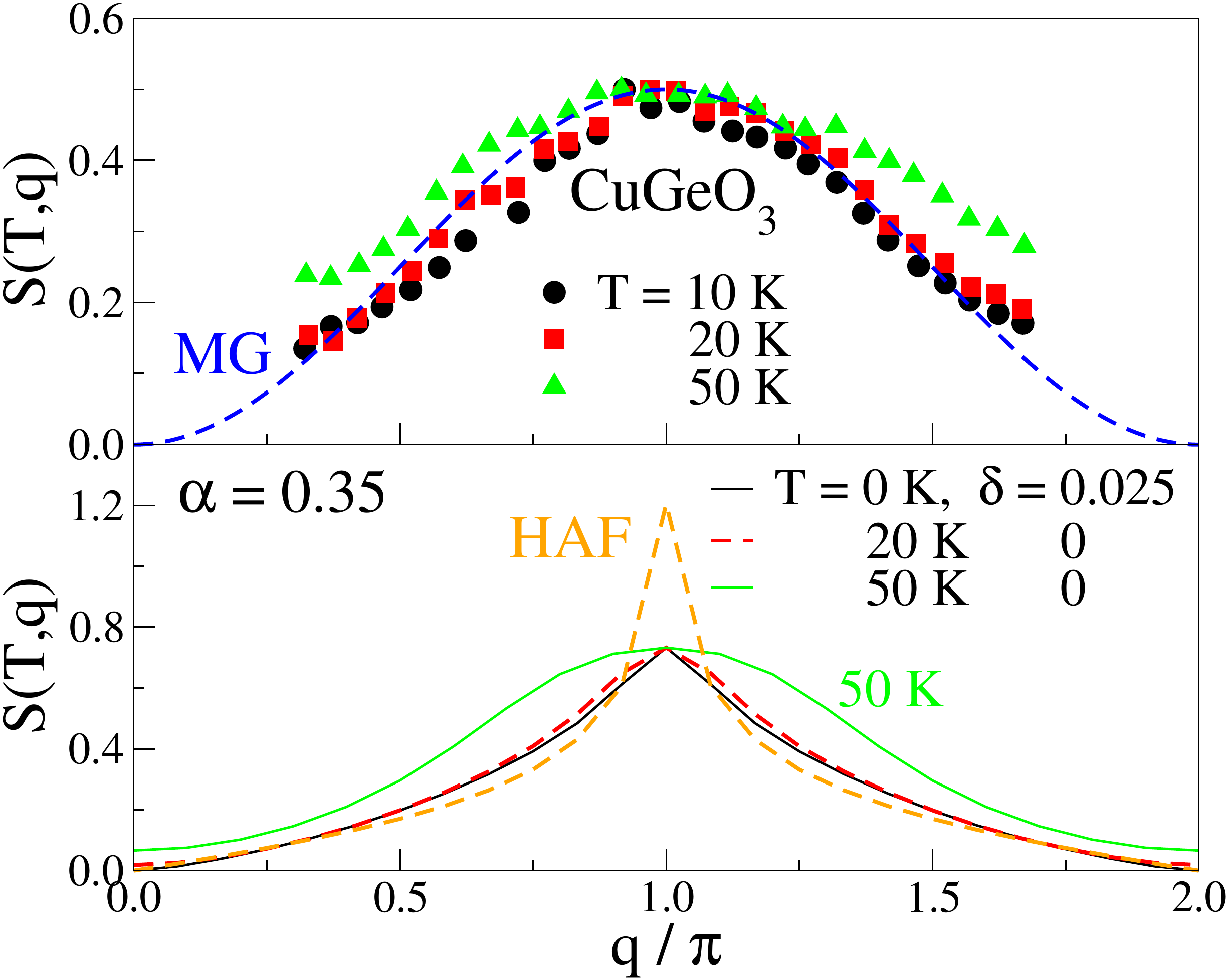}
\caption{\label{fig10}
Upper panel: Static structure factor $S(T,q)$ rescaled from Fig. 2 of Ref.~\onlinecite{arai1996}. The exact $S(q)$ at the MG point is
$(1 - \cos q)/4$ at $T = 0$. Lower panel: Calculated $S(T,q)$ for $N = 24$ spins at $\alpha= 0.35$ and $0.0$. The area under $T = 0$ curves
is $\pi/2$ in both panels. The HAF peak at $q = \pi$ diverges in the thermodynamic limit.
        }
\end{figure}

The calculated $S(T,q)$ in the lower panel of Fig.~\ref{fig10} are for the HAF and for CuGeO$_3$ parameters: $J_1 = 160$ K, $\alpha= 0.35$ and 
$T_{SP} = 14$ K. The HAF structure factor for $N = 24$, $\delta= 0$ is strongly peaked at $q = \pi$ and diverges in 
the thermodynamic limit, but the size dependence elsewhere is small since the area is conserved~\cite{mkumar2015}. The $\alpha= 0.35$ 
curves at $T = 0$ and $20$ K ($0.125$) are for $N = 24$ and $\delta= 0.025$ and $0$, respectively. We obtained explicitly the $T$ 
dependence of the spin correlation functions in $S(q)$. The $T = 0$ and $20$ K curves illustrate 
similar spin correlations at $\delta> 0$, $T = 0$ and $T > 0$, $\delta= 0$, as seen in experiment. Convolution with a broadening 
function in $q$ will be needed to match the observed peaks that depend on resolution in $q$. The $T = 50$ K ($0.313$) line is based on 
$\delta= 0$, $N = 20$ since the thermodynamic limit is reached at lower $T$. We understand the modest broadening at $50$ K by noting that 
$J_1 = 160$ K is large. We conclude that a 1D model with correlated states accounts reasonably well for these INS data.

\section{\label{sec5} Discussion}

Structural changes at Peierls or SP transitions as well as thermal expansion or contraction have been probed by 
elastic X-ray or neutron scattering, as discussed in reviews of widely different classes of quasi-1D crystals~\cite{jerome2004,
pouget2017,special_neural}. We note that 3D changes are always found that in some cases exceed those of the 1D chain. While delicate growth 
conditions leave open the definitive CuGeO$_3$ structure, the largest change below $T_{SP}$ is along the $b$ axis rather than the 
$\sim 1\% $ dimerization of the chain along the $c$ axis~\cite{harris94}. Small displacement $\pm u$ of Cu ions along $c$ is consistent 
with the small calculated $\delta(0) = 0.025$ in the correlated model; they are related by the linear spin-phonon coupling constant. 
Elastic scattering below $T_{SP}$ from superlattice points is due to structural changes, not just dimerization, that are all 
initiated at $T_{SP}$ but do not necessarily vary identically with $T$. Elastic superlattice scattering and coupling to 3D lattices are 
beyond the scope of this paper.

We have applied the hybrid ED/DMRG method to the best characterized SP transitions and to the $J_1-J_2$ model, Eq.~\ref{eq:j1j2}, with frustration
$0 \le \alpha \le 0.50$ and isotropic exchange $J_1$, $J_2=\alpha J_1$ between first and second neighbors. We 
exploit the fact that $\delta(0)$ limits the range of spin correlations at $T = 0$ while finite $T$
limits the range at $\delta= 0$. Internal consistency requires $T_{SP}$ to govern both the stiffness $1/\varepsilon_d$ and dimerization $\delta(T)$.
The relevant system size depends on $T_{SP}/J_1$, about $50$ spins for the transitions modeled. When the thermodynamic limit can be reached, 
the SP transition becomes essentially model exact. On the other hand, the SP instability at $T \sim 0$ is mathematically motivated and 
beyond the hybrid method. The \textit{general} problem is the SP transition at arbitrary $T_{SP}$ while we have modeled \textit{specific} systems with known $T_{SP}$.

Correlated states account quantitatively for the magnetic susceptibility of both crystals. On the theoretical side, $\delta(T)$ of that HAF deviates from 
free fermions or BCS, contrary to previous expectations based on mean field. We place CuGeO$_3$ in the dimer phase with $\alpha= 0.35$ on the basis of $\chi(T)$, 
the specific heat, the ratio $\Delta(T)/\Delta(0)$ of the singlet-triplet gap and INS data that provide an independent determination of $J_1 = 160$ K. 
The first inorganic SP system is also, to the best of our knowledge, the first physical realization of the dimer phase of the $J_1-J_2$ model.

Isotropic exchange between near neighbors is the dominant magnetic interaction that governs the thermodynamics of spin chains. 
However, such 1D models are approximate and incomplete. Approximate because spin-orbit coupling generates corrections to isotropic exchange and $g$ 
factors that are more important in Cu(II) systems than for organic radicals. Incomplete because dipolar interaction between spins are neglected, 
as well as hyperfine interactions with nuclear spins and all interchain interactions. More detailed analysis of specific quasi-1D systems beyond, 
for example, the $J_1-J_2$ model will certainly be needed at low $T$. Neutron~\cite{nishi1994,lussier1996,martin1996,regnault96} and esr~\cite{hori96} 
data indicate $J^\prime \sim J_1/10$ between chains and corrections to 
isotropic exchange, respectively, in CuGeO$_3$. The present results establish that the $J_1-J_2$ model is the proper starting point for finer low-$T$ modeling.  

A static magnetic field $H$ can readily be added to Eq.~\ref{eq:j1j2} as $-g\mu_{B}HS^{Z}$ where $S^Z$ is the total spin component 
along $H$ and $\mu_{B}$ is the Bohr magneton. Since total $S$ is conserved, the energy spectrum $\lbrace E(\delta,N) \rbrace$ of correlated states has resolved Zeeman energies when $H > 0$, and the tensor $g$ may often be taken as a scalar. Multiple studies 
of the SP transition of CuGeO$_3$ in applied fields a few Tesla have been reviewed~\cite{uchinokura2002}. 
The field dependence has been successfully modeled. We 
anticipate at most minor changes on analyzing magnetic field effects using correlated states.

We are computing correlation functions of spin-$1/2$ chains as functions of $T$ and $H$ and separation between 
spins. One goal is to quantify $S(T,q)$, the $T$ dependence of the static structure factor, Eq.~\ref{eq:static_str_fac}, in models with increasing 
frustration $\alpha$. A limitation of the hybrid method became apparent in connection with $S(q,\omega)$. While ED is computed in sectors 
with fixed $q$ and $S^Z$, DMRG is performed in sectors with fixed $S^Z$ up to a cutoff $E_C(\delta,N)$. We can infer $S$ and $q$, but the $S^{Z}=1$ 
states below the cutoff cluster around $q \sim \pi$ is Fig.~\ref{fig9}. The spectrum around $q \sim \pi/2$ that starts at $E>1$ is 
soon above $E_C(\delta,N)$ with increasing system size.

In summary, we have modeled the SP transition of the HAF and $J_1-J_2$ model with $\alpha \le 0.5$ in Eq.~\ref{eq:j1j2} using correlated states. 
The thermodynamic limit of finite chains is reached under conditions that are satisfied by $T_{SP}$ of 
TTF-CuS$_4$C$_4$(CF$_3$)$_4$ and CuGeO$_3$. 
The SP transition depends strongly on frustration $\alpha = J_2/J_1$ because the $\alpha < \alpha_c = 0.2214$ phase is gapless with a nondegenerate ground 
state while $\alpha > \alpha_c$ is gapped with a doubly degenerate ground state.

\begin{acknowledgments}
We thank T. Lorenz for providing us the $\chi(T)$ data.
ZGS thanks D. Huse for several clarifying discussions.
SKS thanks DST-INSPIRE for financial support.
MK thanks DST India for financial support through a Ramanujan fellowship. 
\end{acknowledgments}

\appendix*
\section{}

Choose the Kekul\'e diagram with $N/2$ singlet pairs at sites $2r,2r-1$ as the ground state $\vert G \rangle$ at the MG point. The $J_1-J_2$ model reads

\begin{equation}
\begin{aligned}
        & \qquad \qquad  H (\frac{1}{2},0) =H_0 + \\ 
&\sum_{r=1}^{N/2} \left( {\vec{S}}_{2r} \cdot {\vec{S}}_{2r+1}+\frac{1}{2} \left( {\vec{S}}_{2r-1} \cdot {\vec{S}}_{2r+1} + {\vec{S}}_{2r} \cdot {\vec{S}}_{2r+2}\right) \right).
\label{eq:appen_hamiltonian}
\end{aligned}
\end{equation}
$H_0$ describes isolated dimers $2r$, $2r-1$ with singlet-triplet gap $E=1$. The second term acts on adjacent singlets pairs 
in $\vert G \rangle$. Direct multiplication of spin functions shows that each term annihilates $\vert G \rangle$; 
$E_0 = -3N/8$ is exact.

Let $\vert 2m,2m-1\rangle$ be the product function with a triplet at sites $2m,2m-1$ and singlets at sites $2r,2r-1,r \ne m$. 
The triplet degeneracy under $H_0$ is $N/2$. The second term still annihilates $\vert 2m , 2m-1 \rangle$ when acting on adjacent singlets, 
but not when acting on the triplet and either adjacent singlet. Annihilation requires an out-of-phase linear combination of 
triplets that occurs at $q = \pi/2$ or $3\pi/2$,
\begin{equation}
	\vert T, \pi/2 \rangle = \left( \dfrac{2}{N} \right)^{\frac{1}{2}} \sum_{m=1}^{\frac{N}{2}} (-1)^{m} \vert 2m, 2m-1 \rangle.
\end{equation}
The normalized triplet $\vert T,\pi/2\rangle$ is an exact excited state in the thermodynamic limit. In Eq.~\ref{eq:matelem}, we find 
that $S^Z_{\pi/2} \vert G \rangle = (1/2)\vert T,\pi/2\rangle$. All INS intensity at $q = \pi/2$ or $3\pi/2$ is at $E=1$. 
The other Kekul\'e diagram with singlet pairs $2r,2r+1$ gives the same result.


\begin{thebibliography}{45}%
\makeatletter
\providecommand \@ifxundefined [1]{%
 \@ifx{#1\undefined}
}%
\providecommand \@ifnum [1]{%
 \ifnum #1\expandafter \@firstoftwo
 \else \expandafter \@secondoftwo
 \fi
}%
\providecommand \@ifx [1]{%
 \ifx #1\expandafter \@firstoftwo
 \else \expandafter \@secondoftwo
 \fi
}%
\providecommand \natexlab [1]{#1}%
\providecommand \enquote  [1]{``#1''}%
\providecommand \bibnamefont  [1]{#1}%
\providecommand \bibfnamefont [1]{#1}%
\providecommand \citenamefont [1]{#1}%
\providecommand \href@noop [0]{\@secondoftwo}%
\providecommand \href [0]{\begingroup \@sanitize@url \@href}%
\providecommand \@href[1]{\@@startlink{#1}\@@href}%
\providecommand \@@href[1]{\endgroup#1\@@endlink}%
\providecommand \@sanitize@url [0]{\catcode `\\12\catcode `\$12\catcode
  `\&12\catcode `\#12\catcode `\^12\catcode `\_12\catcode `\%12\relax}%
\providecommand \@@startlink[1]{}%
\providecommand \@@endlink[0]{}%
\providecommand \url  [0]{\begingroup\@sanitize@url \@url }%
\providecommand \@url [1]{\endgroup\@href {#1}{\urlprefix }}%
\providecommand \urlprefix  [0]{URL }%
\providecommand \Eprint [0]{\href }%
\providecommand \doibase [0]{http://dx.doi.org/}%
\providecommand \selectlanguage [0]{\@gobble}%
\providecommand \bibinfo  [0]{\@secondoftwo}%
\providecommand \bibfield  [0]{\@secondoftwo}%
\providecommand \translation [1]{[#1]}%
\providecommand \BibitemOpen [0]{}%
\providecommand \bibitemStop [0]{}%
\providecommand \bibitemNoStop [0]{.\EOS\space}%
\providecommand \EOS [0]{\spacefactor3000\relax}%
\providecommand \BibitemShut  [1]{\csname bibitem#1\endcsname}%
\let\auto@bib@innerbib\@empty
\bibitem [{\citenamefont {Jacobs}\ \emph {et~al.}(1976)\citenamefont {Jacobs},
  \citenamefont {Bray}, \citenamefont {Hart}, \citenamefont {Interrante},
  \citenamefont {Kasper}, \citenamefont {Watkins}, \citenamefont {Prober},\
  and\ \citenamefont {Bonner}}]{jacob1976}%
  \BibitemOpen
  \bibfield  {author} {\bibinfo {author} {\bibfnamefont {I.~S.}\ \bibnamefont
  {Jacobs}}, \bibinfo {author} {\bibfnamefont {J.~W.}\ \bibnamefont {Bray}},
  \bibinfo {author} {\bibfnamefont {H.~R.}\ \bibnamefont {Hart}}, \bibinfo
  {author} {\bibfnamefont {L.~V.}\ \bibnamefont {Interrante}}, \bibinfo
  {author} {\bibfnamefont {J.~S.}\ \bibnamefont {Kasper}}, \bibinfo {author}
  {\bibfnamefont {G.~D.}\ \bibnamefont {Watkins}}, \bibinfo {author}
  {\bibfnamefont {D.~E.}\ \bibnamefont {Prober}}, \ and\ \bibinfo {author}
  {\bibfnamefont {J.~C.}\ \bibnamefont {Bonner}},\ }\href {\doibase
  10.1103/PhysRevB.14.3036} {\bibfield  {journal} {\bibinfo  {journal} {Phys.
  Rev. B}\ }\textbf {\bibinfo {volume} {14}},\ \bibinfo {pages} {3036}
  (\bibinfo {year} {1976})}\BibitemShut {NoStop}%
\bibitem [{\citenamefont {Hase}\ \emph
  {et~al.}(1993{\natexlab{a}})\citenamefont {Hase}, \citenamefont {Terasaki},\
  and\ \citenamefont {Uchinokura}}]{haseprl1993}%
  \BibitemOpen
  \bibfield  {author} {\bibinfo {author} {\bibfnamefont {M.}~\bibnamefont
  {Hase}}, \bibinfo {author} {\bibfnamefont {I.}~\bibnamefont {Terasaki}}, \
  and\ \bibinfo {author} {\bibfnamefont {K.}~\bibnamefont {Uchinokura}},\
  }\href {\doibase 10.1103/PhysRevLett.70.3651} {\bibfield  {journal} {\bibinfo
   {journal} {Phys. Rev. Lett.}\ }\textbf {\bibinfo {volume} {70}},\ \bibinfo
  {pages} {3651} (\bibinfo {year} {1993}{\natexlab{a}})}\BibitemShut {NoStop}%
\bibitem [{\citenamefont {Hase}\ \emph
  {et~al.}(1993{\natexlab{b}})\citenamefont {Hase}, \citenamefont {Terasaki},
  \citenamefont {Uchinokura}, \citenamefont {Tokunaga}, \citenamefont {Miura},\
  and\ \citenamefont {Obara}}]{haseprb1993}%
  \BibitemOpen
  \bibfield  {author} {\bibinfo {author} {\bibfnamefont {M.}~\bibnamefont
  {Hase}}, \bibinfo {author} {\bibfnamefont {I.}~\bibnamefont {Terasaki}},
  \bibinfo {author} {\bibfnamefont {K.}~\bibnamefont {Uchinokura}}, \bibinfo
  {author} {\bibfnamefont {M.}~\bibnamefont {Tokunaga}}, \bibinfo {author}
  {\bibfnamefont {N.}~\bibnamefont {Miura}}, \ and\ \bibinfo {author}
  {\bibfnamefont {H.}~\bibnamefont {Obara}},\ }\href {\doibase
  10.1103/PhysRevB.48.9616} {\bibfield  {journal} {\bibinfo  {journal} {Phys.
  Rev. B}\ }\textbf {\bibinfo {volume} {48}},\ \bibinfo {pages} {9616}
  (\bibinfo {year} {1993}{\natexlab{b}})}\BibitemShut {NoStop}%
\bibitem [{\citenamefont {Riera}\ and\ \citenamefont
  {Dobry}(1995)}]{riera1995}%
  \BibitemOpen
  \bibfield  {author} {\bibinfo {author} {\bibfnamefont {J.}~\bibnamefont
  {Riera}}\ and\ \bibinfo {author} {\bibfnamefont {A.}~\bibnamefont {Dobry}},\
  }\href {\doibase 10.1103/PhysRevB.51.16098} {\bibfield  {journal} {\bibinfo
  {journal} {Phys. Rev. B}\ }\textbf {\bibinfo {volume} {51}},\ \bibinfo
  {pages} {16098} (\bibinfo {year} {1995})}\BibitemShut {NoStop}%
\bibitem [{\citenamefont {Uchinokura}(2002)}]{uchinokura2002}%
  \BibitemOpen
  \bibfield  {author} {\bibinfo {author} {\bibfnamefont {K.}~\bibnamefont
  {Uchinokura}},\ }\href {\doibase 10.1088/0953-8984/14/10/201} {\bibfield
  {journal} {\bibinfo  {journal} {J. Phys.: Condens. Matter}\ }\textbf
  {\bibinfo {volume} {14}},\ \bibinfo {pages} {R195} (\bibinfo {year}
  {2002})}\BibitemShut {NoStop}%
\bibitem [{\citenamefont {Saha}\ \emph {et~al.}(2019)\citenamefont {Saha},
  \citenamefont {Dey}, \citenamefont {Kumar},\ and\ \citenamefont
  {Soos}}]{sudip19}%
  \BibitemOpen
  \bibfield  {author} {\bibinfo {author} {\bibfnamefont {S.~K.}\ \bibnamefont
  {Saha}}, \bibinfo {author} {\bibfnamefont {D.}~\bibnamefont {Dey}}, \bibinfo
  {author} {\bibfnamefont {M.}~\bibnamefont {Kumar}}, \ and\ \bibinfo {author}
  {\bibfnamefont {Z.~G.}\ \bibnamefont {Soos}},\ }\href {\doibase
  10.1103/PhysRevB.99.195144} {\bibfield  {journal} {\bibinfo  {journal} {Phys.
  Rev. B}\ }\textbf {\bibinfo {volume} {99}},\ \bibinfo {pages} {195144}
  (\bibinfo {year} {2019})}\BibitemShut {NoStop}%
\bibitem [{\citenamefont {Bonner}\ and\ \citenamefont
  {Fisher}(1964)}]{bonner64}%
  \BibitemOpen
  \bibfield  {author} {\bibinfo {author} {\bibfnamefont {J.~C.}\ \bibnamefont
  {Bonner}}\ and\ \bibinfo {author} {\bibfnamefont {M.~E.}\ \bibnamefont
  {Fisher}},\ }\href {\doibase 10.1103/PhysRev.135.A640} {\bibfield  {journal}
  {\bibinfo  {journal} {Phys. Rev.}\ }\textbf {\bibinfo {volume} {135}},\
  \bibinfo {pages} {A640} (\bibinfo {year} {1964})}\BibitemShut {NoStop}%
\bibitem [{\citenamefont {Su}\ \emph {et~al.}(1980)\citenamefont {Su},
  \citenamefont {Schrieffer},\ and\ \citenamefont {Heeger}}]{su1980}%
  \BibitemOpen
  \bibfield  {author} {\bibinfo {author} {\bibfnamefont {W.~P.}\ \bibnamefont
  {Su}}, \bibinfo {author} {\bibfnamefont {J.~R.}\ \bibnamefont {Schrieffer}},
  \ and\ \bibinfo {author} {\bibfnamefont {A.~J.}\ \bibnamefont {Heeger}},\
  }\href {\doibase 10.1103/PhysRevB.22.2099} {\bibfield  {journal} {\bibinfo
  {journal} {Phys. Rev. B}\ }\textbf {\bibinfo {volume} {22}},\ \bibinfo
  {pages} {2099} (\bibinfo {year} {1980})}\BibitemShut {NoStop}%
\bibitem [{\citenamefont {Beni}\ and\ \citenamefont {Pincus}(1972)}]{beni1972}%
  \BibitemOpen
  \bibfield  {author} {\bibinfo {author} {\bibfnamefont {G.}~\bibnamefont
  {Beni}}\ and\ \bibinfo {author} {\bibfnamefont {P.}~\bibnamefont {Pincus}},\
  }\href {\doibase 10.1063/1.1678789} {\bibfield  {journal} {\bibinfo
  {journal} {J. Chem. Phys.}\ }\textbf {\bibinfo {volume} {57}},\ \bibinfo
  {pages} {3531} (\bibinfo {year} {1972})}\BibitemShut {NoStop}%
\bibitem [{\citenamefont {Del~Freo}\ \emph {et~al.}(2002)\citenamefont
  {Del~Freo}, \citenamefont {Painelli},\ and\ \citenamefont
  {Soos}}]{soosfreo2002}%
  \BibitemOpen
  \bibfield  {author} {\bibinfo {author} {\bibfnamefont {L.}~\bibnamefont
  {Del~Freo}}, \bibinfo {author} {\bibfnamefont {A.}~\bibnamefont {Painelli}},
  \ and\ \bibinfo {author} {\bibfnamefont {Z.~G.}\ \bibnamefont {Soos}},\
  }\href {\doibase 10.1103/PhysRevLett.89.027402} {\bibfield  {journal}
  {\bibinfo  {journal} {Phys. Rev. Lett.}\ }\textbf {\bibinfo {volume} {89}},\
  \bibinfo {pages} {027402} (\bibinfo {year} {2002})}\BibitemShut {NoStop}%
\bibitem [{\citenamefont {Barnes}\ \emph {et~al.}(1999)\citenamefont {Barnes},
  \citenamefont {Riera},\ and\ \citenamefont {Tennant}}]{barnes99}%
  \BibitemOpen
  \bibfield  {author} {\bibinfo {author} {\bibfnamefont {T.}~\bibnamefont
  {Barnes}}, \bibinfo {author} {\bibfnamefont {J.}~\bibnamefont {Riera}}, \
  and\ \bibinfo {author} {\bibfnamefont {D.~A.}\ \bibnamefont {Tennant}},\
  }\href {\doibase 10.1103/PhysRevB.59.11384} {\bibfield  {journal} {\bibinfo
  {journal} {Phys. Rev. B}\ }\textbf {\bibinfo {volume} {59}},\ \bibinfo
  {pages} {11384} (\bibinfo {year} {1999})}\BibitemShut {NoStop}%
\bibitem [{\citenamefont {Johnston}\ \emph {et~al.}(2000)\citenamefont
  {Johnston}, \citenamefont {Kremer}, \citenamefont {Troyer}, \citenamefont
  {Wang}, \citenamefont {Kl\"umper}, \citenamefont {Bud'ko}, \citenamefont
  {Panchula},\ and\ \citenamefont {Canfield}}]{johnston2000}%
  \BibitemOpen
  \bibfield  {author} {\bibinfo {author} {\bibfnamefont {D.~C.}\ \bibnamefont
  {Johnston}}, \bibinfo {author} {\bibfnamefont {R.~K.}\ \bibnamefont
  {Kremer}}, \bibinfo {author} {\bibfnamefont {M.}~\bibnamefont {Troyer}},
  \bibinfo {author} {\bibfnamefont {X.}~\bibnamefont {Wang}}, \bibinfo {author}
  {\bibfnamefont {A.}~\bibnamefont {Kl\"umper}}, \bibinfo {author}
  {\bibfnamefont {S.~L.}\ \bibnamefont {Bud'ko}}, \bibinfo {author}
  {\bibfnamefont {A.~F.}\ \bibnamefont {Panchula}}, \ and\ \bibinfo {author}
  {\bibfnamefont {P.~C.}\ \bibnamefont {Canfield}},\ }\href {\doibase
  10.1103/PhysRevB.61.9558} {\bibfield  {journal} {\bibinfo  {journal} {Phys.
  Rev. B}\ }\textbf {\bibinfo {volume} {61}},\ \bibinfo {pages} {9558}
  (\bibinfo {year} {2000})}\BibitemShut {NoStop}%
\bibitem [{\citenamefont {Kumar}\ \emph {et~al.}(2007)\citenamefont {Kumar},
  \citenamefont {Ramasesha}, \citenamefont {Sen},\ and\ \citenamefont
  {Soos}}]{mkumar2007}%
  \BibitemOpen
  \bibfield  {author} {\bibinfo {author} {\bibfnamefont {M.}~\bibnamefont
  {Kumar}}, \bibinfo {author} {\bibfnamefont {S.}~\bibnamefont {Ramasesha}},
  \bibinfo {author} {\bibfnamefont {D.}~\bibnamefont {Sen}}, \ and\ \bibinfo
  {author} {\bibfnamefont {Z.~G.}\ \bibnamefont {Soos}},\ }\href {\doibase
  10.1103/PhysRevB.75.052404} {\bibfield  {journal} {\bibinfo  {journal} {Phys.
  Rev. B}\ }\textbf {\bibinfo {volume} {75}},\ \bibinfo {pages} {052404}
  (\bibinfo {year} {2007})}\BibitemShut {NoStop}%
\bibitem [{\citenamefont {Bray}\ \emph {et~al.}(1983)\citenamefont {Bray},
  \citenamefont {Interrante}, \citenamefont {Jacobs},\ and\ \citenamefont
  {Bonner}}]{bray1983}%
  \BibitemOpen
  \bibfield  {author} {\bibinfo {author} {\bibfnamefont {J.~W.}\ \bibnamefont
  {Bray}}, \bibinfo {author} {\bibfnamefont {L.~V.}\ \bibnamefont
  {Interrante}}, \bibinfo {author} {\bibfnamefont {I.~S.}\ \bibnamefont
  {Jacobs}}, \ and\ \bibinfo {author} {\bibfnamefont {J.~C.}\ \bibnamefont
  {Bonner}},\ }\enquote {\bibinfo {title} {The spin-peierls transition},}\ in\
  \href@noop {} {\emph {\bibinfo {booktitle} {Extended Linear Chain
  Compounds}}},\ Vol.~\bibinfo {volume} {3},\ \bibinfo {editor} {edited by\
  \bibinfo {editor} {\bibfnamefont {J.~S.}\ \bibnamefont {Miller}}}\ (\bibinfo
  {publisher} {Plenum Press},\ \bibinfo {address} {New York},\ \bibinfo {year}
  {1983})\ pp.\ \bibinfo {pages} {353--416}\BibitemShut {NoStop}%
\bibitem [{\citenamefont {Majumdar}\ and\ \citenamefont
  {Ghosh}(1969)}]{ckm69b}%
  \BibitemOpen
  \bibfield  {author} {\bibinfo {author} {\bibfnamefont {C.~K.}\ \bibnamefont
  {Majumdar}}\ and\ \bibinfo {author} {\bibfnamefont {D.~K.}\ \bibnamefont
  {Ghosh}},\ }\href {\doibase 10.1063/1.1664979} {\bibfield  {journal}
  {\bibinfo  {journal} {J. Math. Phys.}\ }\textbf {\bibinfo {volume} {10}},\
  \bibinfo {pages} {1399} (\bibinfo {year} {1969})}\BibitemShut {NoStop}%
\bibitem [{\citenamefont {Okamoto}\ and\ \citenamefont
  {Nomura}(1992)}]{nomura1992}%
  \BibitemOpen
  \bibfield  {author} {\bibinfo {author} {\bibfnamefont {K.}~\bibnamefont
  {Okamoto}}\ and\ \bibinfo {author} {\bibfnamefont {K.}~\bibnamefont
  {Nomura}},\ }\href {https://doi.org/10.1016/0375-9601(92)90823-5} {\bibfield
  {journal} {\bibinfo  {journal} {Phys. Lett. A}\ }\textbf {\bibinfo {volume}
  {169}},\ \bibinfo {pages} {433 } (\bibinfo {year} {1992})}\BibitemShut
  {NoStop}%
\bibitem [{\citenamefont {Dey}\ \emph {et~al.}(2016)\citenamefont {Dey},
  \citenamefont {Maiti},\ and\ \citenamefont {Kumar}}]{ddpbc2016}%
  \BibitemOpen
  \bibfield  {author} {\bibinfo {author} {\bibfnamefont {D.}~\bibnamefont
  {Dey}}, \bibinfo {author} {\bibfnamefont {D.}~\bibnamefont {Maiti}}, \ and\
  \bibinfo {author} {\bibfnamefont {M.}~\bibnamefont {Kumar}},\ }\href
  {\doibase 10.4279/pip.080006} {\bibfield  {journal} {\bibinfo  {journal}
  {Papers in Physics}\ }\textbf {\bibinfo {volume} {8}} (\bibinfo {year}
  {2016}),\ 10.4279/pip.080006}\BibitemShut {NoStop}%
\bibitem [{\citenamefont {Saha}\ \emph {et~al.}()\citenamefont {Saha},
  \citenamefont {Kumar},\ and\ \citenamefont {Soos}}]{saha2019spinpeierls}%
  \BibitemOpen
  \bibfield  {author} {\bibinfo {author} {\bibfnamefont {S.~K.}\ \bibnamefont
  {Saha}}, \bibinfo {author} {\bibfnamefont {M.}~\bibnamefont {Kumar}}, \ and\
  \bibinfo {author} {\bibfnamefont {Z.~G.}\ \bibnamefont {Soos}},\ }\href@noop
  {} {}\Eprint {http://arxiv.org/abs/1907.03724} {arXiv:1907.03724
  [cond-mat.str-el]} \BibitemShut {NoStop}%
\bibitem [{\citenamefont {Kumar}\ \emph {et~al.}(2010)\citenamefont {Kumar},
  \citenamefont {Ramasesha},\ and\ \citenamefont {Soos}}]{mkumarbndord2010}%
  \BibitemOpen
  \bibfield  {author} {\bibinfo {author} {\bibfnamefont {M.}~\bibnamefont
  {Kumar}}, \bibinfo {author} {\bibfnamefont {S.}~\bibnamefont {Ramasesha}}, \
  and\ \bibinfo {author} {\bibfnamefont {Z.~G.}\ \bibnamefont {Soos}},\ }\href
  {\doibase 10.1103/PhysRevB.81.054413} {\bibfield  {journal} {\bibinfo
  {journal} {Phys. Rev. B}\ }\textbf {\bibinfo {volume} {81}},\ \bibinfo
  {pages} {054413} (\bibinfo {year} {2010})}\BibitemShut {NoStop}%
\bibitem [{\citenamefont {Cross}\ and\ \citenamefont {Fisher}(1979)}]{cross79}%
  \BibitemOpen
  \bibfield  {author} {\bibinfo {author} {\bibfnamefont {M.~C.}\ \bibnamefont
  {Cross}}\ and\ \bibinfo {author} {\bibfnamefont {D.~S.}\ \bibnamefont
  {Fisher}},\ }\href {\doibase 10.1103/PhysRevB.19.402} {\bibfield  {journal}
  {\bibinfo  {journal} {Phys. Rev. B}\ }\textbf {\bibinfo {volume} {19}},\
  \bibinfo {pages} {402} (\bibinfo {year} {1979})}\BibitemShut {NoStop}%
\bibitem [{\citenamefont {Fabricius}\ \emph {et~al.}(1998)\citenamefont
  {Fabricius}, \citenamefont {Kl\"umper}, \citenamefont {L\"ow}, \citenamefont
  {B\"uchner}, \citenamefont {Lorenz}, \citenamefont {Dhalenne},\ and\
  \citenamefont {Revcolevschi}}]{fabricius1998}%
  \BibitemOpen
  \bibfield  {author} {\bibinfo {author} {\bibfnamefont {K.}~\bibnamefont
  {Fabricius}}, \bibinfo {author} {\bibfnamefont {A.}~\bibnamefont
  {Kl\"umper}}, \bibinfo {author} {\bibfnamefont {U.}~\bibnamefont {L\"ow}},
  \bibinfo {author} {\bibfnamefont {B.}~\bibnamefont {B\"uchner}}, \bibinfo
  {author} {\bibfnamefont {T.}~\bibnamefont {Lorenz}}, \bibinfo {author}
  {\bibfnamefont {G.}~\bibnamefont {Dhalenne}}, \ and\ \bibinfo {author}
  {\bibfnamefont {A.}~\bibnamefont {Revcolevschi}},\ }\href {\doibase
  10.1103/PhysRevB.57.1102} {\bibfield  {journal} {\bibinfo  {journal} {Phys.
  Rev. B}\ }\textbf {\bibinfo {volume} {57}},\ \bibinfo {pages} {1102}
  (\bibinfo {year} {1998})}\BibitemShut {NoStop}%
\bibitem [{\citenamefont {Hidaka}\ \emph {et~al.}(1997)\citenamefont {Hidaka},
  \citenamefont {Hatae}, \citenamefont {Yamada}, \citenamefont {Nishi},\ and\
  \citenamefont {Akimitsu}}]{hidaka1997}%
  \BibitemOpen
  \bibfield  {author} {\bibinfo {author} {\bibfnamefont {M.}~\bibnamefont
  {Hidaka}}, \bibinfo {author} {\bibfnamefont {M.}~\bibnamefont {Hatae}},
  \bibinfo {author} {\bibfnamefont {I.}~\bibnamefont {Yamada}}, \bibinfo
  {author} {\bibfnamefont {M.}~\bibnamefont {Nishi}}, \ and\ \bibinfo {author}
  {\bibfnamefont {J.}~\bibnamefont {Akimitsu}},\ }\href {\doibase
  10.1088/0953-8984/9/4/003} {\bibfield  {journal} {\bibinfo  {journal} {J.
  Phys.: Condens. Matter}\ }\textbf {\bibinfo {volume} {9}},\ \bibinfo {pages}
  {809} (\bibinfo {year} {1997})}\BibitemShut {NoStop}%
\bibitem [{\citenamefont {Castilla}\ \emph {et~al.}(1995)\citenamefont
  {Castilla}, \citenamefont {Chakravarty},\ and\ \citenamefont
  {Emery}}]{castilla1995}%
  \BibitemOpen
  \bibfield  {author} {\bibinfo {author} {\bibfnamefont {G.}~\bibnamefont
  {Castilla}}, \bibinfo {author} {\bibfnamefont {S.}~\bibnamefont
  {Chakravarty}}, \ and\ \bibinfo {author} {\bibfnamefont {V.~J.}\ \bibnamefont
  {Emery}},\ }\href {\doibase 10.1103/PhysRevLett.75.1823} {\bibfield
  {journal} {\bibinfo  {journal} {Phys. Rev. Lett.}\ }\textbf {\bibinfo
  {volume} {75}},\ \bibinfo {pages} {1823} (\bibinfo {year}
  {1995})}\BibitemShut {NoStop}%
\bibitem [{\citenamefont {Bouzerar}\ \emph {et~al.}(1999)\citenamefont
  {Bouzerar}, \citenamefont {Legeza},\ and\ \citenamefont
  {Ziman}}]{bouzerar1999}%
  \BibitemOpen
  \bibfield  {author} {\bibinfo {author} {\bibfnamefont {G.}~\bibnamefont
  {Bouzerar}}, \bibinfo {author} {\bibfnamefont {O.}~\bibnamefont {Legeza}}, \
  and\ \bibinfo {author} {\bibfnamefont {T.}~\bibnamefont {Ziman}},\ }\href
  {\doibase 10.1103/PhysRevB.60.15278} {\bibfield  {journal} {\bibinfo
  {journal} {Phys. Rev. B}\ }\textbf {\bibinfo {volume} {60}},\ \bibinfo
  {pages} {15278} (\bibinfo {year} {1999})}\BibitemShut {NoStop}%
\bibitem [{\citenamefont {Lorenz}\ \emph {et~al.}(1996)\citenamefont {Lorenz},
  \citenamefont {Ammerahl}, \citenamefont {Ziemes}, \citenamefont {B\"uchner},
  \citenamefont {Revcolevschi},\ and\ \citenamefont {Dhalenne}}]{lorenz1996}%
  \BibitemOpen
  \bibfield  {author} {\bibinfo {author} {\bibfnamefont {T.}~\bibnamefont
  {Lorenz}}, \bibinfo {author} {\bibfnamefont {U.}~\bibnamefont {Ammerahl}},
  \bibinfo {author} {\bibfnamefont {R.}~\bibnamefont {Ziemes}}, \bibinfo
  {author} {\bibfnamefont {B.}~\bibnamefont {B\"uchner}}, \bibinfo {author}
  {\bibfnamefont {A.}~\bibnamefont {Revcolevschi}}, \ and\ \bibinfo {author}
  {\bibfnamefont {G.}~\bibnamefont {Dhalenne}},\ }\href {\doibase
  10.1103/PhysRevB.54.R15610} {\bibfield  {journal} {\bibinfo  {journal} {Phys.
  Rev. B}\ }\textbf {\bibinfo {volume} {54}},\ \bibinfo {pages} {R15610}
  (\bibinfo {year} {1996})}\BibitemShut {NoStop}%
\bibitem [{\citenamefont {Liu}\ \emph {et~al.}(1995)\citenamefont {Liu},
  \citenamefont {Wosnitza}, \citenamefont {L\"ohneysen},\ and\ \citenamefont
  {Kremer}}]{liu1995}%
  \BibitemOpen
  \bibfield  {author} {\bibinfo {author} {\bibfnamefont {X.}~\bibnamefont
  {Liu}}, \bibinfo {author} {\bibfnamefont {J.}~\bibnamefont {Wosnitza}},
  \bibinfo {author} {\bibfnamefont {H.}~\bibnamefont {L\"ohneysen}}, \ and\
  \bibinfo {author} {\bibfnamefont {R.}~\bibnamefont {Kremer}},\ }\href
  {\doibase 10.1007/BF01324521} {\bibfield  {journal} {\bibinfo  {journal} {Z.
  Phys. B}\ }\textbf {\bibinfo {volume} {98}},\ \bibinfo {pages} {163}
  (\bibinfo {year} {1995})}\BibitemShut {NoStop}%
\bibitem [{\citenamefont {Wei}\ \emph {et~al.}(1977)\citenamefont {Wei},
  \citenamefont {Heeger}, \citenamefont {Salamon},\ and\ \citenamefont
  {Delker}}]{wei1977}%
  \BibitemOpen
  \bibfield  {author} {\bibinfo {author} {\bibfnamefont {T.}~\bibnamefont
  {Wei}}, \bibinfo {author} {\bibfnamefont {A.~J.}\ \bibnamefont {Heeger}},
  \bibinfo {author} {\bibfnamefont {M.~B.}\ \bibnamefont {Salamon}}, \ and\
  \bibinfo {author} {\bibfnamefont {G.~E.}\ \bibnamefont {Delker}},\ }\href
  {\doibase 10.1016/0038-1098(77)90041-2} {\bibfield  {journal} {\bibinfo
  {journal} {Solid State Commun.}\ }\textbf {\bibinfo {volume} {21}},\ \bibinfo
  {pages} {595} (\bibinfo {year} {1977})}\BibitemShut {NoStop}%
\bibitem [{\citenamefont {Nishi}\ \emph {et~al.}(1994)\citenamefont {Nishi},
  \citenamefont {Fujita},\ and\ \citenamefont {Akimitsu}}]{nishi1994}%
  \BibitemOpen
  \bibfield  {author} {\bibinfo {author} {\bibfnamefont {M.}~\bibnamefont
  {Nishi}}, \bibinfo {author} {\bibfnamefont {O.}~\bibnamefont {Fujita}}, \
  and\ \bibinfo {author} {\bibfnamefont {J.}~\bibnamefont {Akimitsu}},\ }\href
  {\doibase 10.1103/PhysRevB.50.6508} {\bibfield  {journal} {\bibinfo
  {journal} {Phys. Rev. B}\ }\textbf {\bibinfo {volume} {50}},\ \bibinfo
  {pages} {6508} (\bibinfo {year} {1994})}\BibitemShut {NoStop}%
\bibitem [{\citenamefont {Lussier}\ \emph {et~al.}(1996)\citenamefont
  {Lussier}, \citenamefont {Coad}, \citenamefont {McMorrow},\ and\
  \citenamefont {Paul}}]{lussier1996}%
  \BibitemOpen
  \bibfield  {author} {\bibinfo {author} {\bibfnamefont {J.-G.}\ \bibnamefont
  {Lussier}}, \bibinfo {author} {\bibfnamefont {S.~M.}\ \bibnamefont {Coad}},
  \bibinfo {author} {\bibfnamefont {D.~F.}\ \bibnamefont {McMorrow}}, \ and\
  \bibinfo {author} {\bibfnamefont {D.~M.}\ \bibnamefont {Paul}},\ }\href
  {\doibase 10.1088/0953-8984/8/4/003} {\bibfield  {journal} {\bibinfo
  {journal} {J. Phys.: Condens. Matter}\ }\textbf {\bibinfo {volume} {8}},\
  \bibinfo {pages} {L59} (\bibinfo {year} {1996})}\BibitemShut {NoStop}%
\bibitem [{\citenamefont {Martin}\ \emph {et~al.}(1996)\citenamefont {Martin},
  \citenamefont {Shirane}, \citenamefont {Fujii}, \citenamefont {Nishi},
  \citenamefont {Fujita}, \citenamefont {Akimitsu}, \citenamefont {Hase},\ and\
  \citenamefont {Uchinokura}}]{martin1996}%
  \BibitemOpen
  \bibfield  {author} {\bibinfo {author} {\bibfnamefont {M.~C.}\ \bibnamefont
  {Martin}}, \bibinfo {author} {\bibfnamefont {G.}~\bibnamefont {Shirane}},
  \bibinfo {author} {\bibfnamefont {Y.}~\bibnamefont {Fujii}}, \bibinfo
  {author} {\bibfnamefont {M.}~\bibnamefont {Nishi}}, \bibinfo {author}
  {\bibfnamefont {O.}~\bibnamefont {Fujita}}, \bibinfo {author} {\bibfnamefont
  {J.}~\bibnamefont {Akimitsu}}, \bibinfo {author} {\bibfnamefont
  {M.}~\bibnamefont {Hase}}, \ and\ \bibinfo {author} {\bibfnamefont
  {K.}~\bibnamefont {Uchinokura}},\ }\href {\doibase
  10.1103/PhysRevB.53.R14713} {\bibfield  {journal} {\bibinfo  {journal} {Phys.
  Rev. B}\ }\textbf {\bibinfo {volume} {53}},\ \bibinfo {pages} {R14713}
  (\bibinfo {year} {1996})}\BibitemShut {NoStop}%
\bibitem [{\citenamefont {Regnault}\ \emph {et~al.}(1996)\citenamefont
  {Regnault}, \citenamefont {A\"{\i}n}, \citenamefont {Hennion}, \citenamefont
  {Dhalenne},\ and\ \citenamefont {Revcolevschi}}]{regnault96}%
  \BibitemOpen
  \bibfield  {author} {\bibinfo {author} {\bibfnamefont {L.~P.}\ \bibnamefont
  {Regnault}}, \bibinfo {author} {\bibfnamefont {M.}~\bibnamefont {A\"{\i}n}},
  \bibinfo {author} {\bibfnamefont {B.}~\bibnamefont {Hennion}}, \bibinfo
  {author} {\bibfnamefont {G.}~\bibnamefont {Dhalenne}}, \ and\ \bibinfo
  {author} {\bibfnamefont {A.}~\bibnamefont {Revcolevschi}},\ }\href {\doibase
  10.1103/PhysRevB.53.5579} {\bibfield  {journal} {\bibinfo  {journal} {Phys.
  Rev. B}\ }\textbf {\bibinfo {volume} {53}},\ \bibinfo {pages} {5579}
  (\bibinfo {year} {1996})}\BibitemShut {NoStop}%
\bibitem [{\citenamefont {M\"uller}\ \emph {et~al.}(1981)\citenamefont
  {M\"uller}, \citenamefont {Thomas}, \citenamefont {Beck},\ and\ \citenamefont
  {Bonner}}]{muller1981}%
  \BibitemOpen
  \bibfield  {author} {\bibinfo {author} {\bibfnamefont {G.}~\bibnamefont
  {M\"uller}}, \bibinfo {author} {\bibfnamefont {H.}~\bibnamefont {Thomas}},
  \bibinfo {author} {\bibfnamefont {H.}~\bibnamefont {Beck}}, \ and\ \bibinfo
  {author} {\bibfnamefont {J.~C.}\ \bibnamefont {Bonner}},\ }\href {\doibase
  10.1103/PhysRevB.24.1429} {\bibfield  {journal} {\bibinfo  {journal} {Phys.
  Rev. B}\ }\textbf {\bibinfo {volume} {24}},\ \bibinfo {pages} {1429}
  (\bibinfo {year} {1981})}\BibitemShut {NoStop}%
\bibitem [{\citenamefont {Bethe}(1931)}]{bethe31}%
  \BibitemOpen
  \bibfield  {author} {\bibinfo {author} {\bibfnamefont {H.}~\bibnamefont
  {Bethe}},\ }\href {\doibase 10.1007/BF01341708} {\bibfield  {journal}
  {\bibinfo  {journal} {Z. Phys.}\ }\textbf {\bibinfo {volume} {71}},\ \bibinfo
  {pages} {205} (\bibinfo {year} {1931})}\BibitemShut {NoStop}%
\bibitem [{\citenamefont {{Hulth\'en}}(1938)}]{hulthen38}%
  \BibitemOpen
  \bibfield  {author} {\bibinfo {author} {\bibfnamefont {L.}~\bibnamefont
  {{Hulth\'en}}},\ }\href@noop {} {\bibfield  {journal} {\bibinfo  {journal}
  {{Ark. Mat. Astron. Fys.}}\ }\textbf {\bibinfo {volume} {26}},\ \bibinfo
  {pages} {106} (\bibinfo {year} {1938})}\BibitemShut {NoStop}%
\bibitem [{\citenamefont {Faddeev}\ and\ \citenamefont
  {Takhtajan}(1981)}]{faddeev81}%
  \BibitemOpen
  \bibfield  {author} {\bibinfo {author} {\bibfnamefont {L.}~\bibnamefont
  {Faddeev}}\ and\ \bibinfo {author} {\bibfnamefont {L.}~\bibnamefont
  {Takhtajan}},\ }\href@noop {} {\bibfield  {journal} {\bibinfo  {journal}
  {Physics Letters A}\ }\textbf {\bibinfo {volume} {85}},\ \bibinfo {pages}
  {375 } (\bibinfo {year} {1981})}\BibitemShut {NoStop}%
\bibitem [{\citenamefont {des Cloizeaux}\ and\ \citenamefont
  {Pearson}(1962)}]{pearson1962}%
  \BibitemOpen
  \bibfield  {author} {\bibinfo {author} {\bibfnamefont {J.}~\bibnamefont {des
  Cloizeaux}}\ and\ \bibinfo {author} {\bibfnamefont {J.~J.}\ \bibnamefont
  {Pearson}},\ }\href {\doibase 10.1103/PhysRev.128.2131} {\bibfield  {journal}
  {\bibinfo  {journal} {Phys. Rev.}\ }\textbf {\bibinfo {volume} {128}},\
  \bibinfo {pages} {2131} (\bibinfo {year} {1962})}\BibitemShut {NoStop}%
\bibitem [{\citenamefont {Karbach}\ \emph {et~al.}(1997)\citenamefont
  {Karbach}, \citenamefont {M\"uller}, \citenamefont {Bougourzi}, \citenamefont
  {Fledderjohann},\ and\ \citenamefont {M\"utter}}]{karbach97}%
  \BibitemOpen
  \bibfield  {author} {\bibinfo {author} {\bibfnamefont {M.}~\bibnamefont
  {Karbach}}, \bibinfo {author} {\bibfnamefont {G.}~\bibnamefont {M\"uller}},
  \bibinfo {author} {\bibfnamefont {A.~H.}\ \bibnamefont {Bougourzi}}, \bibinfo
  {author} {\bibfnamefont {A.}~\bibnamefont {Fledderjohann}}, \ and\ \bibinfo
  {author} {\bibfnamefont {K.-H.}\ \bibnamefont {M\"utter}},\ }\href {\doibase
  10.1103/PhysRevB.55.12510} {\bibfield  {journal} {\bibinfo  {journal} {Phys.
  Rev. B}\ }\textbf {\bibinfo {volume} {55}},\ \bibinfo {pages} {12510}
  (\bibinfo {year} {1997})}\BibitemShut {NoStop}%
\bibitem [{\citenamefont {Mourigal}\ \emph {et~al.}(2013)\citenamefont
  {Mourigal}, \citenamefont {Enderle}, \citenamefont {Kl\"opperpieper},
  \citenamefont {Caux}, \citenamefont {Stunault},\ and\ \citenamefont
  {R\o{}nnow}}]{mourigal2013}%
  \BibitemOpen
  \bibfield  {author} {\bibinfo {author} {\bibfnamefont {M.}~\bibnamefont
  {Mourigal}}, \bibinfo {author} {\bibfnamefont {M.}~\bibnamefont {Enderle}},
  \bibinfo {author} {\bibfnamefont {A.}~\bibnamefont {Kl\"opperpieper}},
  \bibinfo {author} {\bibfnamefont {J.-S.}\ \bibnamefont {Caux}}, \bibinfo
  {author} {\bibfnamefont {A.}~\bibnamefont {Stunault}}, \ and\ \bibinfo
  {author} {\bibfnamefont {H.~M.}\ \bibnamefont {R\o{}nnow}},\ }\href {\doibase
  10.1038/nphys2652} {\bibfield  {journal} {\bibinfo  {journal} {Nature
  Physics}\ }\textbf {\bibinfo {volume} {9}},\ \bibinfo {pages} {435} (\bibinfo
  {year} {2013})}\BibitemShut {NoStop}%
\bibitem [{\citenamefont {Arai}\ \emph {et~al.}(1996)\citenamefont {Arai},
  \citenamefont {Fujita}, \citenamefont {Motokawa}, \citenamefont {Akimitsu},\
  and\ \citenamefont {Bennington}}]{arai1996}%
  \BibitemOpen
  \bibfield  {author} {\bibinfo {author} {\bibfnamefont {M.}~\bibnamefont
  {Arai}}, \bibinfo {author} {\bibfnamefont {M.}~\bibnamefont {Fujita}},
  \bibinfo {author} {\bibfnamefont {M.}~\bibnamefont {Motokawa}}, \bibinfo
  {author} {\bibfnamefont {J.}~\bibnamefont {Akimitsu}}, \ and\ \bibinfo
  {author} {\bibfnamefont {S.~M.}\ \bibnamefont {Bennington}},\ }\href
  {\doibase 10.1103/PhysRevLett.77.3649} {\bibfield  {journal} {\bibinfo
  {journal} {Phys. Rev. Lett.}\ }\textbf {\bibinfo {volume} {77}},\ \bibinfo
  {pages} {3649} (\bibinfo {year} {1996})}\BibitemShut {NoStop}%
\bibitem [{\citenamefont {Kumar}\ \emph {et~al.}(2015)\citenamefont {Kumar},
  \citenamefont {Parvej},\ and\ \citenamefont {Soos}}]{mkumar2015}%
  \BibitemOpen
  \bibfield  {author} {\bibinfo {author} {\bibfnamefont {M.}~\bibnamefont
  {Kumar}}, \bibinfo {author} {\bibfnamefont {A.}~\bibnamefont {Parvej}}, \
  and\ \bibinfo {author} {\bibfnamefont {Z.~G.}\ \bibnamefont {Soos}},\ }\href
  {\doibase 10.1088/0953-8984/27/31/316001} {\bibfield  {journal} {\bibinfo
  {journal} {Journal of Physics: Condensed Matter}\ }\textbf {\bibinfo {volume}
  {27}},\ \bibinfo {pages} {316001} (\bibinfo {year} {2015})}\BibitemShut
  {NoStop}%
\bibitem [{\citenamefont {J\'erome}(2004)}]{jerome2004}%
  \BibitemOpen
  \bibfield  {author} {\bibinfo {author} {\bibfnamefont {D.}~\bibnamefont
  {J\'erome}},\ }\href {https://doi.org/10.1021/cr030652g} {\bibfield
  {journal} {\bibinfo  {journal} {Chem. Rev.}\ }\textbf {\bibinfo {volume}
  {104}},\ \bibinfo {pages} {5565} (\bibinfo {year} {2004})}\BibitemShut
  {NoStop}%
\bibitem [{\citenamefont {Pouget}\ \emph {et~al.}(2017)\citenamefont {Pouget},
  \citenamefont {Foury-Leylekian},\ and\ \citenamefont {Almeida}}]{pouget2017}%
  \BibitemOpen
  \bibfield  {author} {\bibinfo {author} {\bibfnamefont {J.-P.}\ \bibnamefont
  {Pouget}}, \bibinfo {author} {\bibfnamefont {P.}~\bibnamefont
  {Foury-Leylekian}}, \ and\ \bibinfo {author} {\bibfnamefont {M.}~\bibnamefont
  {Almeida}},\ }\href {https://doi.org/10.3390/magnetochemistry3010013}
  {\bibfield  {journal} {\bibinfo  {journal} {Magnetochemistry}\ }\textbf
  {\bibinfo {volume} {3}} (\bibinfo {year} {2017})}\BibitemShut {NoStop}%
\bibitem [{\citenamefont {Painelli}\ and\ \citenamefont
  {Girlando}(2017)}]{special_neural}%
  \BibitemOpen
  \bibinfo {editor} {\bibfnamefont {A.}~\bibnamefont {Painelli}}\ and\ \bibinfo
  {editor} {\bibfnamefont {A.}~\bibnamefont {Girlando}},\ eds.,\ \href
  {https://www.mdpi.com/journal/crystals/special_issues/The_neutral_ionic_phase_transition#}
  {\emph {\bibinfo {title} {The Neutral-Ionic Phase Transition}}}\ (\bibinfo
  {publisher} {Crystals 7},\ \bibinfo {year} {2017})\BibitemShut {NoStop}%
\bibitem [{\citenamefont {Harris}\ \emph {et~al.}(1994)\citenamefont {Harris},
  \citenamefont {Feng}, \citenamefont {Birgeneau}, \citenamefont {Hirota},
  \citenamefont {Kakurai}, \citenamefont {Lorenzo}, \citenamefont {Shirane},
  \citenamefont {Hase}, \citenamefont {Uchinokura}, \citenamefont {Kojima},
  \citenamefont {Tanaka},\ and\ \citenamefont {Shibuya}}]{harris94}%
  \BibitemOpen
  \bibfield  {author} {\bibinfo {author} {\bibfnamefont {Q.~J.}\ \bibnamefont
  {Harris}}, \bibinfo {author} {\bibfnamefont {Q.}~\bibnamefont {Feng}},
  \bibinfo {author} {\bibfnamefont {R.~J.}\ \bibnamefont {Birgeneau}}, \bibinfo
  {author} {\bibfnamefont {K.}~\bibnamefont {Hirota}}, \bibinfo {author}
  {\bibfnamefont {K.}~\bibnamefont {Kakurai}}, \bibinfo {author} {\bibfnamefont
  {J.~E.}\ \bibnamefont {Lorenzo}}, \bibinfo {author} {\bibfnamefont
  {G.}~\bibnamefont {Shirane}}, \bibinfo {author} {\bibfnamefont
  {M.}~\bibnamefont {Hase}}, \bibinfo {author} {\bibfnamefont {K.}~\bibnamefont
  {Uchinokura}}, \bibinfo {author} {\bibfnamefont {H.}~\bibnamefont {Kojima}},
  \bibinfo {author} {\bibfnamefont {I.}~\bibnamefont {Tanaka}}, \ and\ \bibinfo
  {author} {\bibfnamefont {Y.}~\bibnamefont {Shibuya}},\ }\href {\doibase
  10.1103/PhysRevB.50.12606} {\bibfield  {journal} {\bibinfo  {journal} {Phys.
  Rev. B}\ }\textbf {\bibinfo {volume} {50}},\ \bibinfo {pages} {12606}
  (\bibinfo {year} {1994})}\BibitemShut {NoStop}%
\bibitem [{\citenamefont {Honda}\ \emph {et~al.}(1996)\citenamefont {Honda},
  \citenamefont {Shibata}, \citenamefont {Kindo}, \citenamefont {Sugai},
  \citenamefont {Takeuchi},\ and\ \citenamefont {Hori}}]{hori96}%
  \BibitemOpen
  \bibfield  {author} {\bibinfo {author} {\bibfnamefont {M.}~\bibnamefont
  {Honda}}, \bibinfo {author} {\bibfnamefont {T.}~\bibnamefont {Shibata}},
  \bibinfo {author} {\bibfnamefont {K.}~\bibnamefont {Kindo}}, \bibinfo
  {author} {\bibfnamefont {S.}~\bibnamefont {Sugai}}, \bibinfo {author}
  {\bibfnamefont {T.}~\bibnamefont {Takeuchi}}, \ and\ \bibinfo {author}
  {\bibfnamefont {H.}~\bibnamefont {Hori}},\ }\href {\doibase
  10.1143/JPSJ.65.691} {\bibfield  {journal} {\bibinfo  {journal} {J. Phys.
  Soc. Jpn.}\ }\textbf {\bibinfo {volume} {65}},\ \bibinfo {pages} {691}
  (\bibinfo {year} {1996})}\BibitemShut {NoStop}%
\end{thebibliography}

%

\end{document}